\documentclass{caps}
\usepackage{psfig,latexsym,longtable,lscape}

\def\etal{ {\it et al. }}

\def\ergs{\hbox{erg s$^{-1}$ }}

\def\ergcms{\hbox{erg cm$^{-2}$ s$^{-1}$ }}

\def\msun{$M_{\odot}$}

\def\p0{\phantom{0}}

\def\it{\sl}

\DeclareRobustCommand{\ion}[2]{%
\relax\ifmmode
\ifx\testbx\f@series
{\mathbf{#1\,\mathsc{#2}}}\else
{\mathrm{#1\,\mathsc{#2}}}\fi
\else\textup{#1\,{\mdseries\textsc{#2}}}%
\fi}

\begin{document}

\pagenumbering{roman}
%\tableofcontents
\maketitle
\cleardoublepage

\pagenumbering{arabic}

\setcounter{chapter}{10}

\setcounter{table}{0}

\author[G. Fabbiano \& N. White]{G. Fabbiano$^1$ \& N. E. White$^2$ \\\ 
$^1$Harvard-Smithsonian Center for Astrophysics\\
60 Garden St., Cambridge MA 02138, USA\\
$^2$NASA-GSFC\\ Code 660, Greenbelt MD 20771, USA}

\chapter{Compact Stellar X-ray Sources in Normal Galaxies}

%%%\vskip -1.5cm
%%%\today

\section{Introduction \label{intro}}

In the 1995 {\it X-ray Binaries} book edited by Lewin, van Paradijs and van den Heuvel,
the chapter on {\it Normal galaxies and their X-ray binary populations} (Fabbiano 1995)
began with the claim that ``X-ray binaries are an important 
component of the X-ray emission of galaxies. Therefore the knowledge 
gathered from the study of Galactic X-ray sources can be used to 
interpret X-ray observations of external galaxies. Conversely,
observations of external galaxies can provide us with uniform samples
of X-ray binaries, in a variety of different environments. '' This statement was
based mostly on the {\it Einstein
Observatory}  survey of normal galaxies (e.g., Fabbiano 1989; Fabbiano, Kim \& Trinchieri 1992).
Those results have been borne out by later work, yet at the time the claim took a certain 
leap of faith. Now, nearly a decade later, the sensitive sub-arcsecond 
spectrally-resolved images of galaxies from {\it Chandra}
(Weisskopf {\it et al.} 2000), complemented by the {\it XMM-Newton} (Jansen {\it et al.} 2001) data 
for the nearest galaxies (angular resolution of {\it XMM-Newton} is $\sim 15"$),
have made strikingly true what was then largely just wishful anticipation.

While a substantial body of {\it {\it ROSAT}} and {\it ASCA} observations exists, 
which was not included in the 1995 Chapter, the revolutionary quality of the {\it Chandra} 
(and to a more
limited degree of {\it XMM-Newton}) data is such that the present review will be based
on these most recent results.

In this Chapter we first discuss the emerging awareness of X-ray ($0.1-10$~keV 
band, approximately) stellar populations 
in spiral galaxies: we focus on four well studied galaxies (M31, M81, M83 and
M101), and we then discuss the effect of recent widespread star formation on the 
luminosity functions of the X-ray emitting populations (Section~\ref{pops}). 
We then review the body of
observational evidence on the ultraluminous
X-ray sources ($L_X > 10^{39} ~\rm ergs~s^{-1}$), 
that are associated with active/recent star formation (Section~\ref{ulx};
see the Chapter by King, in this book,
for a review of theoretical work on this subject; see also the chapter
by McClintock \& Remillard on black hole binaries). We follow with a review of the
X-ray population properties of old stellar systems (E and S0 galaxies; Section~\ref{ell}).
We then discuss the results of correlation analyses of the integrated
galaxy emission (Section~\ref{corr}), and we conclude with
a look at the X-ray evolution of galaxies going back into the deep universe (Section~\ref{deep}).

\section{X-ray binary (XRB) Populations  in Spiral Galaxies \label{pops}}

Because of their proximity, nearby spiral galaxies is where the early work on
extra-galactic XRB populations 
begun (see Fabbiano 1995). For the same reason, these are the galaxies where
the deepest samples of sources have been acquired with {\it Chandra}
and {\it XMM-Newton}. 
Here we will discuss first the
recent work done on M31, which, not surprisingly, is the galaxy that has been studied
in most detail. We will then review the results on M81, M83, and M101, to provide 
examples of the XRB populations in a wider variety of spirals. We conclude
this section with a summary of the work on actively star-forming galaxies.
We note that this field is evolving rapidly, with an increasing number of galaxies
being surveyed and with the sensitivity limit being pushed to fainter fluxes,
with ever deeper {\it Chandra} observations. 

\subsection{M31}

Being at a distance of only $\sim 700$~kpc, M31 (NGC~224, the Andromeda nebula) is
the spiral (Sb) galaxy closest to us. M31 has been observed 
by virtually all the X-ray observatories since {\it Uhuru}, the first X-ray
satellite (for a history of the X-ray observations of galaxies, see
Fabbiano \& Kessler 2001). Starting with the {\it
Einstein Observatory} and following on with {\it ROSAT}, 
M31 has been the prime target for systematic studies of
a population of extragalactic XRBs, and for comparisons with our own Galactic
XRBs (e.g., Long \& Van~Speybroeck 1983; Trinchieri \& Fabbiano 1991; 
Primini, Forman \& Jones 1993; Supper {\it et al.} 1997, 2001). 
{\it Chandra} and {\it XMM-Newton} observations, both by themselves and in 
combination, are providing new insight on the characteristics of the XRB population
of M31. With its subarcsecond resolution, {\it Chandra} is unique in resolving dense
source regions, such as the circum-nuclear region of M31, 
and detecting faint sources (Garcia {\it et al.} 2000). Given the proximity of M31 and 
the relatively low density of luminous XRBs, {\it XMM-Newton} provides valuable data
on the XRB population of this galaxy,
if one excludes the centermost crowded core (Shirey {\it et al.} 2001).

\smallskip
\noindent{\bf Source variability and counterparts - }
Multiple observations of the same fields with these two observatories (and comparison with
previous observations) have confirmed the general source variability characteristic of XRBs.
{\it XMM-Newton} work, following the first statement of source variability
(Osborne {\it et al.} 2001), includes detailed studies of interesting luminous sources. 
Trudolyubov, Borozdin \& Priedhorsky (2001) report the discovery of three transient
sources, with maximum X-ray emission in the $10^{37} ~\rm ergs~s^{-1}$ range: a candidate
low-mass black-hole binary, a source with a long ($>$1 year)
outburst, and a supersoft transient. Trudolyubov {\it et al.} (2002b) report
an 83\% modulation with a 2.78~hr period in the X-ray source associated with the
globular cluster (GC) Bo~158. Comparison with earlier {\it XMM-Newton} observations and with
the {\it {\it ROSAT}} PSPC data, allows these authors to conclude that the modulation is anticorrelated
with the source flux, suggesting perhaps a larger less obscured emission region in
high state. This source resembles Galactic `dip' XRBs, and could be an accreting neutron star. 
Its period suggest a highly compact system (separation $\sim 10^{11} \rm cm$). 

Widespread source variability is evident from {\it Chandra} observations, 
both from a  47~ks HRC study of the bulge
(Kaaret 2002), from a set of eight {\it Chandra} ACIS observations of the  central 17'$\times$17'
taken between 1999 and 2001 (Kong {\it et al.} 2002), and from a 2.5 years 17 epochs survey
with the {\it Chandra} HRC (Williams {\it et al.} 2003), which also includes the data from 
Kaaret (2002).

Kong {\it et al.}  find 204 sources, including nine supersoft sources, 
with a detection limit of $\sim 2 \times 10^{35} ~\rm ergs~s^{-1}$. 
This detection limit is 5 times fainter than that of the {\it {\it ROSAT}} HRI catalog (Primini,
Forman \& Jones 1993), which lists only 77 sources in the surveyed area.
They report 22 globular cluster (GC) identifications, 2 supernova remnants, and 9 planetary nebulae associations.
By comparing the different individual data sets,
they establish that 50\% of the sources vary on timescales of months, and 13 are transients.
The spectra of the most luminous sources can be fitted with power-laws with $\Gamma \sim 1.8$,
and, of these, 12 show coordinated flux and spectral variability. Two 
sources exhibit harder spectra with increasing count rate, reminiscent of
Galactic Z sources (e.g. Hasinger \& van der Klis 1989). All these characteristics
point to an XRB population similar to that of the Milky Way.
The HRC survey (Williams {\it et al.} 2003) reports fluxes and light curves for 173 sources,
and finds variability in 25\% of the sources; 17 of these sources are transients, and two of these
are identified with variable {\it HST WFPC2} U band counterparts. One of these two sources is
also a transient in the optical and has global properties suggesting a $\sim 10~M_{\odot}$
black hole X-ray nova with a period $\geq 9$~days. Williams {\it et al.} (2003) determine
that at any given time there are $1.9 \pm 1.3$ active X-ray transients in M31, and from here
they infer that the ratio of neutron star to black hole LMXBs in M31 is $\sim 1$, comparable to that 
in the Galaxy. 

\smallskip
\noindent{\bf Globular Cluster (GC) sources  - }
The recent X-ray populations studies of M31 with {\it Chandra} and {\it XMM-Newton}
demonstrate the importance of large area surveys of the entire galaxian system.
A targeted study of GCs with three {\it Chandra} fields at large galactocentric radii
(Di Stefano {\it et al.} 2002) revives the old suggestion (Long \& Van Speybroeck 1983) that
the M31 GC sources are more  X-ray luminous than Galactic GC sources. This hypothesis
had been dismissed with the {\it {\it ROSAT}} M31 survey (Supper {\it et al.} 1997), which however covered
only the central 34' of M31. Di~Stefano {\it et al.} (2002) find that in their fields the most
luminous sources are associated with GCs. They detect 28 GCs sources, 15 of which are
new detections: 1/3 of these sources have $L_X(0.5-7~{\rm keV}) > 10^{37} ~\rm ergs~s^{-1}$;
1/10 of the sources have $L_X(0.5-7~{\rm keV}) >10^{38} ~\rm ergs~s^{-1}$ . 
The X-ray luminosity function (XLF) of the M31 GC sources differs from the Galactic GC XLF, by
both having a larger number of sources, and by extending  a decade higher in X-ray 
luminosity (the most luminous M31 GC is Bo~375 with $L_X > 2 \times 10^{38} ~\rm ergs~s^{-1}$;
compare with Milky Way GCs, that emit less than $10^{37} ~\rm ergs~s^{-1}$).

\smallskip
\noindent{\bf Supersoft sources (SSS) - }
SSS are very soft X-ray sources, with most of the emission below 1~keV, and spectra that
can be fitted with black body temperatures of $\leq 100$~eV (see Chapter by Kahabka in this volume).
SSS were first discovered in M31 with {\it ROSAT} (Supper {\it et al. } 1997).
As noted above, Kong {\it et al.} (2002) reported nine SSS in their {\it Chandra} observations of M31.
Recent work by Di~Stefano {\it et al.} (2003) reports 33 SSSs in the same fields surveyed
for GCs by Di Stefano {\it et al.} (2002), of which only two were known since the {\it ROSAT} times.
Two SSSs are identified with symbiotic stars and two with supernova remnants, 
but the bulk are likely to be 
supersoft XRBs. These sources are highly variable, and may be classified in two spectral groups:
sources with kT$\leq 100$~eV, and other sources with harder emission, up to kT$\sim 300$~eV.
Sixteen of them (on average the most luminous) cluster in the bulge, 
others are found in both the disk and the halo of M31.
Di~Stefano {\it et al.} (2003) point out that some of these sources are detected 
with luminosities well below $10^{37} \rm ergs~s^{-1}$, the
luminosity of a 0.6~$M_{\odot}$ white dwarf steadily burning hydrogen, and are therefore likely
to be lower mass white dwarfs or luminous cataclysmic variables.

\smallskip
\noindent{\bf The bulge - }
 The XLFs of the global core population [Kaaret 2002 ({\it Chandra} HRC); 
Kong {\it et al.} 2002 ({\it Chandra} ACIS); Trudolyubov {\it et al.} 2002a  ({\it XMM-Newton})] all are in general
agreement with each other and with the {\it Einstein} (Trinchieri \& Fabbiano 1991) and {\it {\it ROSAT}}
studies (Primini, Forman \& Jones 1993). 
However, because of the resolution and sensitivity of {\it Chandra},
both  Kong {\it et al.} (2002) and Kaaret (2002) can can look at the bulge source population
in greater detail than ever before.

Kong {\it et al.} divide the detected sources in three groups,
based on their galactocentric position: inner bulge (2`$\times$2'), outer bulge (8'$\times$8',
excluding the inner bulge sources), and disk (17'$\times$17', excluding the two bulge regions). 
When considering the entire bulge population, these authors
find a general low luminosity break of the XLF at $\sim 2 \times 10^{37} ~\rm ergs~s^{-1}$, 
in agreement with Trudolyubov {\it et al.} (2002a).
However, they also find that the break appears to shift to lower luminosities
with decreasing galactocentric radius, going from $0.18 \pm 0.08 \times 10^{37} ~\rm ergs~s^{-1}$ in
the inner bulge to $2.10 \pm 0.39 \times 10^{37} ~\rm ergs~s^{-1}$ in the outermost `disk'
region. They note that if the breaks mark episodes of star formation, the more recent
of these events must have occurred at larger radii. The slopes of the XLFs also 
vary ($0.67 \pm 0.08$ in the center, $1.86 \pm 0.40$ in the outermost region), but this 
trend is the opposite of that expected from progressively young populations, where more
luminous, short lived sources, may be found (see e.g. Kilgard {\it et al.} 2002; Zezas \& Fabbiano 2002;
Section~\ref{xlf}).
Kong {\it et al.} suggest that the XRB populations of the central regions of M31 may instead all be old
(see Trudolyubov {\it et al.} 2002a),
with the shifts of the break resulting from the inclusion of new classes of fainter sources 
in the inner regions, rather than from a disappearance of the most luminous sources.

\begin{figure*}[htbp]
  \vbox{\hskip -\leftskip
  \psfig{figure=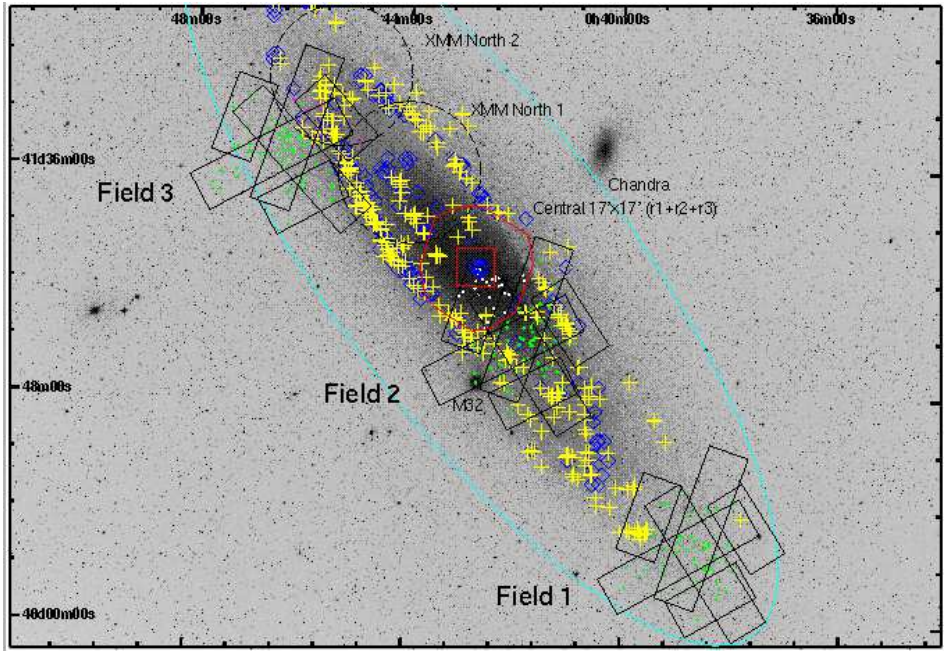,width=12.5cm,angle=0.0}
  }\par
  %,bbllx=1.5cm,bblly=13.7cm,bburx=15.5cm,bbury=25.7cm,clip=}}\par
  \caption{\label{fig:m31fields}
  Regions of M31 observed with {\it Chandra}
  and {\it XMM-Newton}. Dots are detected {\it Chandra} sources; yellow crosses
  and blue diamonds identify supernova remnants and OB associations in the field (not
  X-ray sources), respectively 
  (from Kong {\it et al.} 2003). }
\end{figure*}

Kaaret (2002) contributes to the debate on the nature of the inner bulge sources by
investigating their spatial distribution. 
He shows that the the number of X-ray sources detected in the centermost
regions of the bulge ($< 100''$) is in excess of what would be expected on the basis
of the radial distribution of the optical surface brightness, and suggests that this
result may be consistent with a GC origin for the LMXBs.

\smallskip
\noindent{\bf X-ray source populations in different galaxian fields - }
With the increased rate of papers on M31, resulting from the {\it XMM-Newton}
and {\it Chandra} surveys of this galaxy, we are now realizing that the 
X-ray source population of M31 is more varied than
previously thought, and that there are correlations between the properties
of the X-ray sources and those of the stellar field to which they belong.

In contrast with previous reports (e.g. Trinchieri
\& Fabbiano 1991; Kong {\it et al.} 2002), Trudolyubov {\it et al.} (2002a), by using a larger
definition for the radius of the bulge (15'), with {\it XMM-Newton} observations conclude that,
although the XLFs of bulge and disk sources have a similar cumulative slope (-1.3), disk
sources are all fainter than $L_X < 2 \times 10^{37} ~\rm ergs~s^{-1}$, while bulge sources
can have luminosities as high as $L_X \sim 10^{38} ~\rm ergs~s^{-1}$. 
They suggest that 
the most luminous sources are associated with the older stellar population, 
as in the Milky Way (Grimm, Gilfanov \& Sunyaev 2002).
However, the fields studied by Trudolyubov {\it et al.} (2002a)  do not include the areas surveyed
by Di~Stefano {\it et al.} (2002), where the most luminous GC sources are found 
(see Fig.~\ref{fig:m31fields}).

\begin{figure*}[htbp]
  \vbox{\hskip -\leftskip
  \psfig{figure=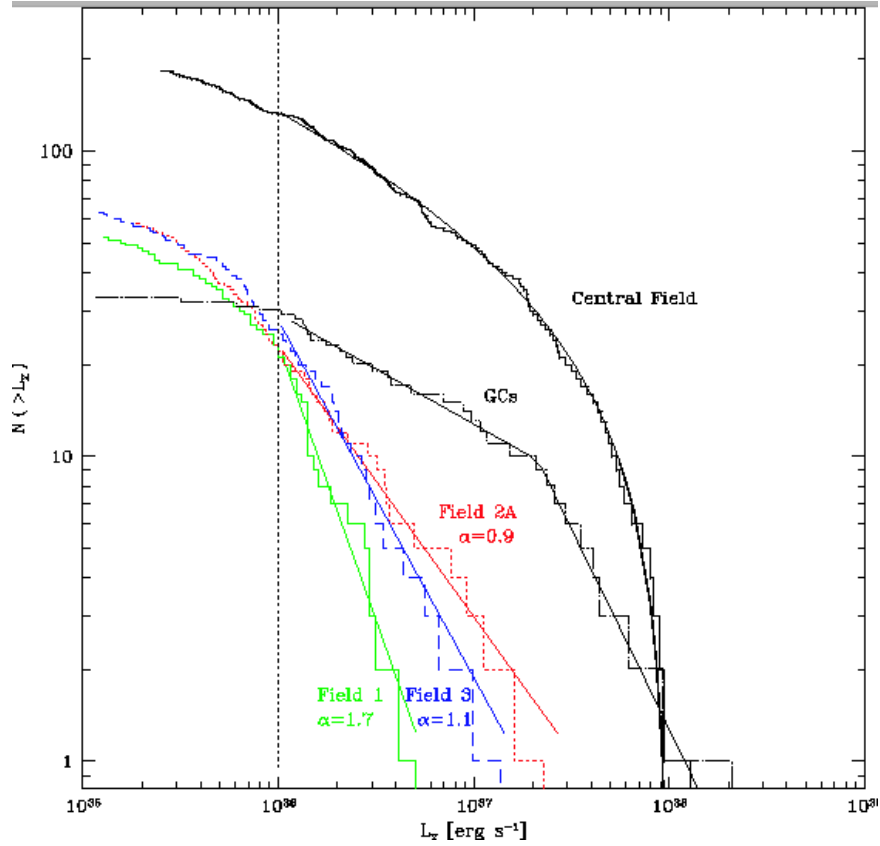,width=11.5cm,angle=0.0}
  }\par
  %,bbllx=1.5cm,bblly=13.7cm,bburx=15.5cm,bbury=25.7cm,clip=}}\par
  \caption{\label{fig:m31XLF}
  Cumulative XLFs and  best-fit power-laws from different fields of M31 (Kong {\it et al.} 2003).}
\end{figure*}

A {\it Chandra ACIS} study of XLFs from different regions of M31 (Fig.~\ref{fig:m31fields}; 
Kong {\it et al.} 2003), uses a 
follow-up of the Di~Stefano {\it et al.} (2002) survey. 
The results (Fig.~\ref{fig:m31XLF}) show that the sources in the 
central 17'$\times$17' region are overall more luminous than those from the outer fields
(as noticed by Trudolyubov {\it et al.} 2002a), but only if one removes the GC population, which
appears to have a relatively more numerous high luminosity component than the central
sources. The slopes of the XLFs of the external fields also vary, and there is an indication
that these differences are related to variations in the stellar populations of the different
fields: Field~1, which has the steepest 
slope (cumulative -1.7$^{+0.34}_{-0. 15}$) and also the lowest density of X-ray sources, 
does not appear to have a large young
population of stars; Field~2, with the largest X-ray source population and the flattest XLF
slope (cumulative -0.9) is in the region with the youngest stellar population. This slope is 
the closest to that ($0.63 \pm 0.13$) derived by Grimm, Gilfanov \& Sunyaev (2002) 
for the high mass X-ray binaries (HMXBs) in the Galaxy; Field~3, with
an intermediate XLF slope instead does not appear to cover a large stellar population.
The overall integrated slope is instead similar to that found by Grimm {\it et al.} for the Galactic
low-mass X-ray binary (LMXB) population, 
suggesting that these sources dominate the X-ray emission of M31.

Williams {\it et al.} (2003), using the {\it Chandra HRC} survey of M31, 
distinguish between a roughly radially symmetric
bulge population (within a 7' radius) and a field population, outside this inner region.
They report different XLFs for bulge and disk sources, with a flatter broken power-law 
representing well the disk distribution. Their survey has a wider (although shallower)
coverage of the entire M31 galaxy, than the Trudolyubov {\it et al.} (2202a) work, and
covers also the southern half of the disk, where the
X-ray sources are significantly more luminous than in the northern disk,
surveyed with {\it XMM-Newton} by Trudolyubov {\it et al.}.

The Trudolyubov {\it et al.} (2202a), Kong {\it et al.} (2003), and Williams {\it et al .} 
(2003) papers are illuminating
in demonstrating the variability of the XLF in different regions, and in pointing out
how a good spatial sampling and supporting multi-wavelength information, are
needed to get a complete picture of the XRB population of M31. 

\subsection{M81}

As discussed in Fabbiano (1995), M81 (NGC~3031) is a nearby ($ 3.6$~Mpc, Freedman {\it et al.} 1994)
Sb galaxy optically similar to M31; however, in X-rays
it displays a significantly more luminous population of individual sources (even discounting
the nuclear AGN). 
To get a feel of the progress in sensitivity of X-ray telescopes in the last
$\sim 20$ years, it is interesting to compare the {\it Einstein} observations of M81,
where 9 extra-nuclear sources with  $L_X \geq 2 \times  10^{38} ~\rm ergs~s^{-1}$
were detected (Fabbiano 1988; total $\sim 35$~ks exposure time), with
the {\it {\it ROSAT}} results  that led to detection 
of 26 extra-nuclear sources with $L_X > 10^{37} ~\rm ergs~s^{-1}$ 
(Immler \& Wang 2001; 177~ks - HRI, 101~ks - PSPC), and finally with the {\it Chandra} results: 
124 sources detected within the optical
$D_{25}$ isophote to a limiting luminosity of $\sim 3 \times 10^{36} ~\rm ergs~s^{-1}$ in $\sim$50~ks
(Swartz {\it et al.} 2003). 

The {\it Chandra} results show that 88\% of the non-nuclear emission is
resolved into individual sources. The brightest of these sources have luminosities exceeding the 
Eddington luminosity for a spherically accreting neutron star (see Fabbiano 1995), i.e.
they are among the sources dubbed `Ultraluminous X-ray Sources' (ULX; see Section~\ref{ulx}).
Of the 66 sources that lie within {\it Hubble Space Telescope (HST)} fields, 
34 have potential counterparts (but 20$\pm$4
chance coincidences are expected). Five sources are coincident with supernova remnants in the
spiral arms (including the well studied SN 1993J), but
one of them (the ULX X-6) is identified with a XRB, based on it X-ray spectrum.
Only four potential GC identifications are found. For one of the M81 sources, Ghosh {\it et al.} (2001)
report a 10-year {\it {\it ROSAT}}-{\it Chandra} X-ray transient light curve.

Nine of the sources found in the {\it Chandra} observation of M81 are supersoft (SSS; Swartz {\it et al.} 2002),
with $L_X(0.2-2.0~{\rm keV})$ in the range of $> 2 \times 10^{36} - 3 \times 10^{38} ~\rm ergs~s^{-1}$,
and a blackbody emission temperature of 40-80~eV. The fraction of SSS is consistent with the expected values, based
on the Galaxy and M31. Four sources are in the bulge and five in the disk; of the latter, four are on
the spiral arms. With the exception of the most luminous of these systems, which has a bolometric
luminosity $L_{bol} \sim 1.5 \times 10^{39} ~\rm ergs~s^{-1}$, and will be discussed in Section~\ref{ulx},
all these sources are consistent with the nuclear-burning accreting white dwarf picture of SSS
(van den Heuvel {\it et al.} 1992; see the Chapter by Kahabka in this book). 
The SSS associated with the spiral arms tend to have higher emission temperatures, suggesting more massive
white dwarf counterparts, which would result from relatively massive stars in a relatively younger
stellar population.

\begin{figure*}[htbp]
  \vbox{\hskip -\leftskip
  \psfig{figure=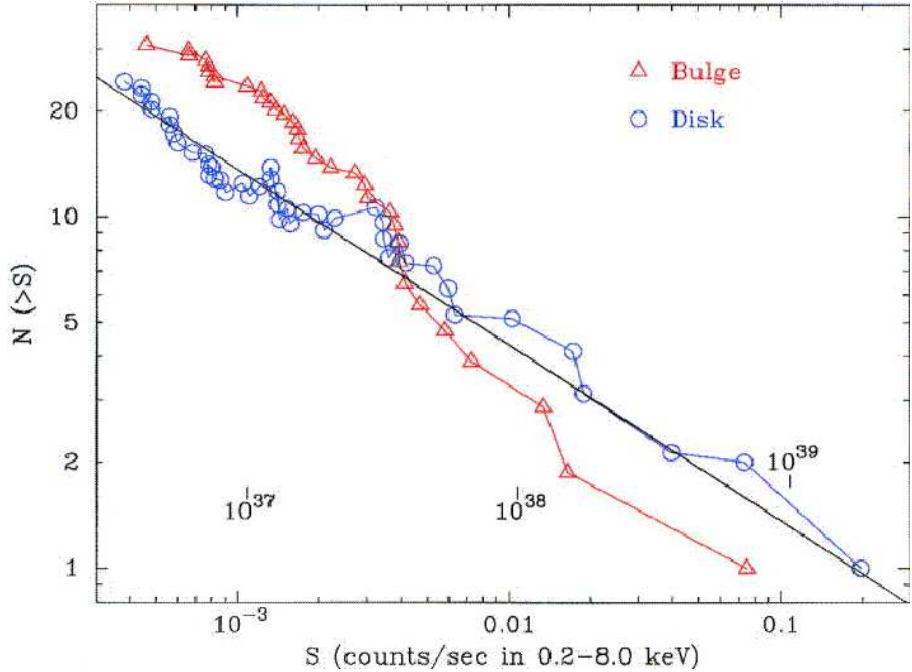,width=12.0cm,angle=0.0}
  }\par
  %,bbllx=1.5cm,bblly=13.7cm,bburx=15.5cm,bbury=25.7cm,clip=}}\par
  \caption{\label{fig:m81XLF}
  Bulge and Disk XLFs for M81. The straight line is the best-fit power-law to the
  disk XLF  (Tennant {\it et al.} 2001).}
\end{figure*}

\begin{figure*}[htbp]
  \vbox{\hskip -\leftskip
  \psfig{figure=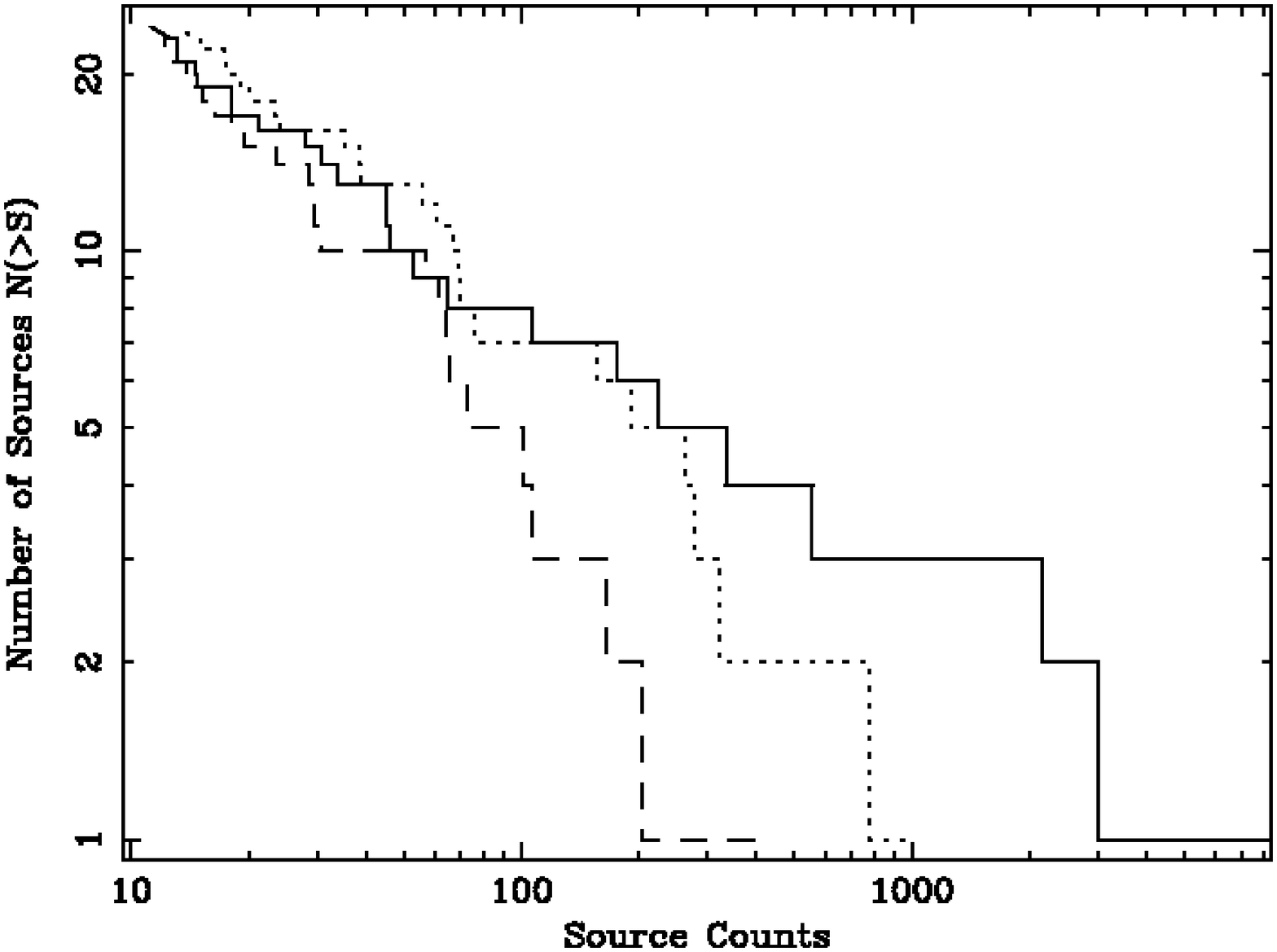,width=12.0cm,angle=270.0}}\par
  %,bbllx=1.5cm,bblly=13.7cm,bburx=15.5cm,bbury=25.7cm,clip=}}\par
  \caption{\label{fig:m81XLFdisk}
  Steepening XLFs of disk sources of M81, at increasing distance from the spiral arms (solid line;
  Swartz {\it et al.} 2003).}
\end{figure*}

The first report of XLF studies in M81 (Tennant {\it et al.} 2001; Fig.\ref{fig:m81XLF}) showed 
dramatic differences
in the XLFs of bulge and disk sources. While the XLF of the bulge is reminiscent of the bulge
of M31, with a relatively steep power-law flattening at $L_X(0.2-8.0~{\rm keV}) < 4 \times
10^{37} ~\rm ergs~s^{-1}$, the XLF of the disk follows a uninterrupted shallow power law 
(cumulative slope -0.50). 

The subsequent more complete study of Swartz {\it et al.} (2003) 
confirms the break in the bulge XLF and suggests that it may be due to an aging $\sim 400$~Myr
old population of LMXBs. The extrapolation of this XLF to lower luminosities can only explain 
10\% of the unresolved bulge emission, which, however, has the same spatial distribution as
the detected bulge sources: besides some gaseous emission, this may suggest an undetected
steepening of the XLF due to a yet fainter older
population of sources in the central regions. The disk population has different XLFs, depending
on the source distance from the spiral arms (Fig.~\ref{fig:m81XLFdisk}): 
in particular, the very luminous ($> 10^{38}~\rm ergs~s^{-1}$)
sources responsible for the flat power law are all concentrated on the arms; 
a break at high luminosities appears when spiral arm sources are excluded. Swartz {\it et al.}
(2003) suggest that these most luminous sources
are likely to be very young XRBs resulting from the star formation stimulated by the spiral density waves.

\subsection{M83 and M101 }

M83 (NGC~5236) and M101 (NGC~5457) are both face-on Sc galaxies. M83 is likely to be a 
member of the Centaurus group, with a distance of $\sim 4$~Mpc (de Vaucouleurs {\it et al.} 1991);
M101 is  more distant ($\sim 7$~Mpc; Stetson {\it et al.} 1998), but still in the nearby universe.

\begin{figure*}[htbp]
  \vbox{\hskip -\leftskip
  \psfig{figure=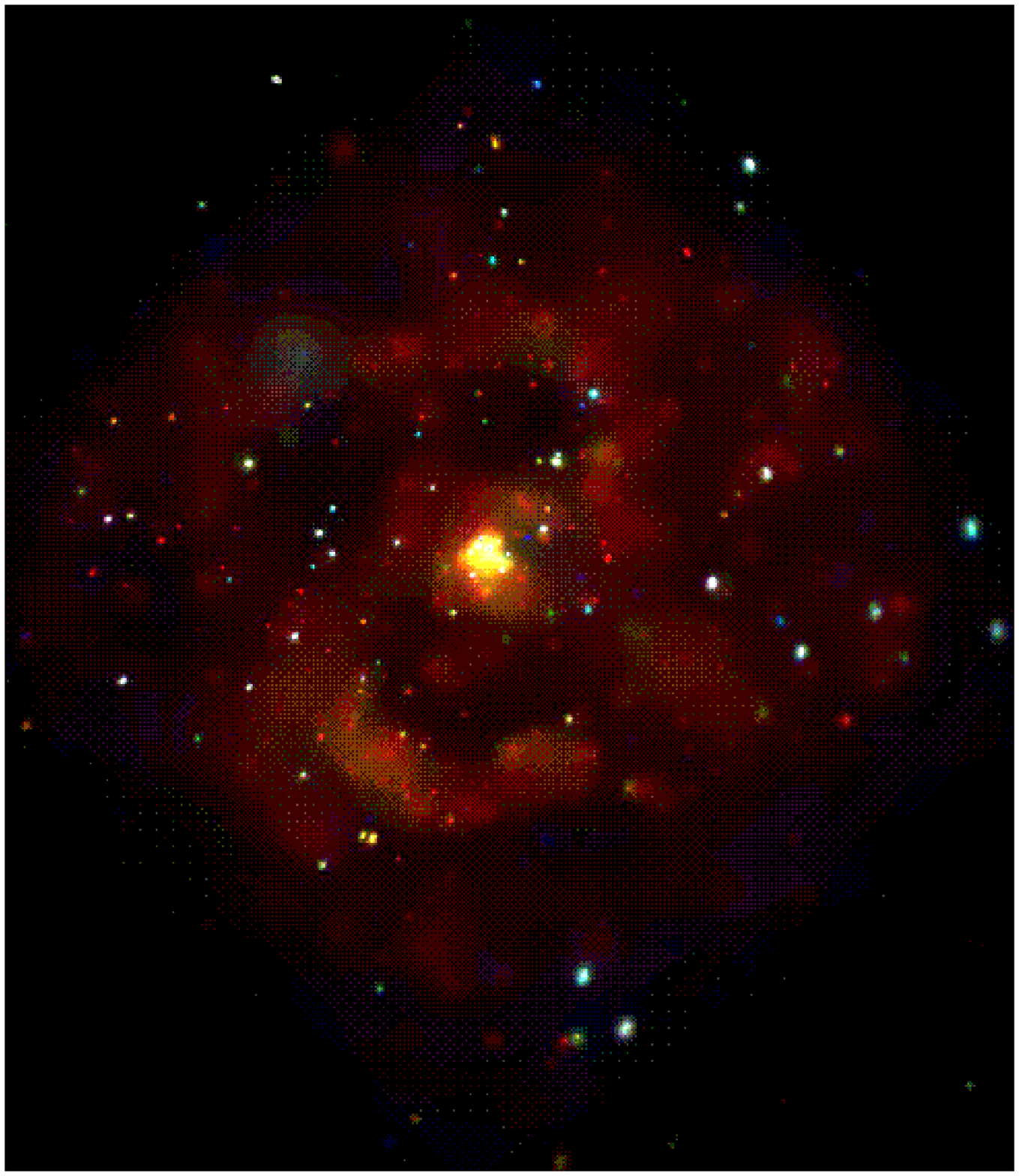,width=12.5cm,angle=0.0}
  }\par
  %,bbllx=1.5cm,bblly=13.7cm,bburx=15.5cm,bbury=25.7cm,clip=}}\par
  \caption{\label{fig:m83}
  M83 as seen with {\it Chandra}.
  Note the population of point-like sources and the softs diffuse emission
  (possibly from hot ISM), associated with the spiral arms  
  (from http://chandra.cfa.harvard.edu/photo/2003/1154/index.html).}
\end{figure*}

M83 is a grand design, barred spiral, with a starburst nucleus. Is has been observed 
extensively in the pre-{\it Chandra} era, but here we discuss only the {\it Chandra}
observations, that are the most relevant for the study of the X-ray source population.
M83 was observed with {\it Chandra} ACIS-S3 for $\sim50$~ks (Fig.~\ref{fig:m83}).
Soria \& Wu (2002) detect 81 sources in these data, of which 18 had been 
detected previously with {\it {\it ROSAT}}; 15 sources are resolved in the previously 
confused nuclear region, which has the highest source density. 
The XLF of the  sources in the nuclear-bar region, where a young stellar population
is likely to prevale, follows a fairly flat unbroken power-law (cumulative slope -0.8).
The XLF of the disk sources is instead steeper (slope -1.3), with a break at $\sim
6 \times 10^{37} ~\rm ergs~s^{-1}$, becoming flatter at the lower luminosities.
This behaviour is reminiscent
of the XLFs of the bulges of M31 and M81, and suggests an older XRB population.

In M101, 110 sources (27 of which are expected to be backround AGN) were detected
in a 98~ks {\it Chandra} ACIS-S3 observation, with a limiting luminosity of
$10^{36} ~\rm ergs~s^{-1}$ (Pence, Snowden \& Mukai 2001).
The sources cluster along the spiral arms, and, interestingly, sources in the 
interarm regions tend to have X-ray colors compatible with AGNs.
Twelve sources are spatially coincident with supernova remnants, but, based on
their variability, two of them are identified with XRBs. Eight other luminous
sources exhibit variability in the {\it {\it Chandra} } data, and two more are found variable
by comparison with previous {\it {\it ROSAT}} observations. Ten sources are supersoft,
and a correlation between black-body temperature and total source luminosity
is suggested by the data. The XLF of the M101 sources can be modelled with
a power-law (cumulative slope -0.8) in the $10^{36} -10^{38} ~\rm ergs~s^{-1}$ 
range.

In summary, with {\it Chandra},  X-ray source population studies are 
finally coming of age. The sub-arcsecond resolution of the {\it Chandra} 
mirrors (Van Speybroeck {\it et al.} 1997) allows both the separation of discrete 
sources from surrounding diffuse emission and the detection of much fainter 
sources than previously possible.

The  XLFs of sources in a given system 
reflect the formation, evolution, and physical properties 
of the X-ray source population. These differences are evident in different
regions of M31, M81 and M83.
Comparison of the XLFs of nearby galaxies (and components thereof)
with the XLFs of more distant systems provides a general coherent
picture, pointing to steeper XLFs in older stellar populations (relative
lack of very luminous sources).
The XLFs of E and S0 galaxies have cumulative slopes in the range
-1.0 to -2.0 (see Section~\ref{exlf}), generally consistent with those of the
bulges of M31 and M81. These slopes are significantly steeper
than those of sources associated with younger stellar fields in M31, M81,
and M83. A recent study of 32 nearby galaxies extracted from the {\it Chandra}
archive (Colbert {\it et al.} 2003) confirms this basic difference between
XLFs of old and younger stellar populations, finding cumulative slopes of
$\sim 1.4$ and $\sim 0.6-0.8$ for elliptical and spiral galaxies respectively.

\subsection{XRBs in Actively Starforming Galaxies \label{xlf}}

\begin{figure*}[htbp]
  \vbox{\hskip -\leftskip
  \psfig{figure=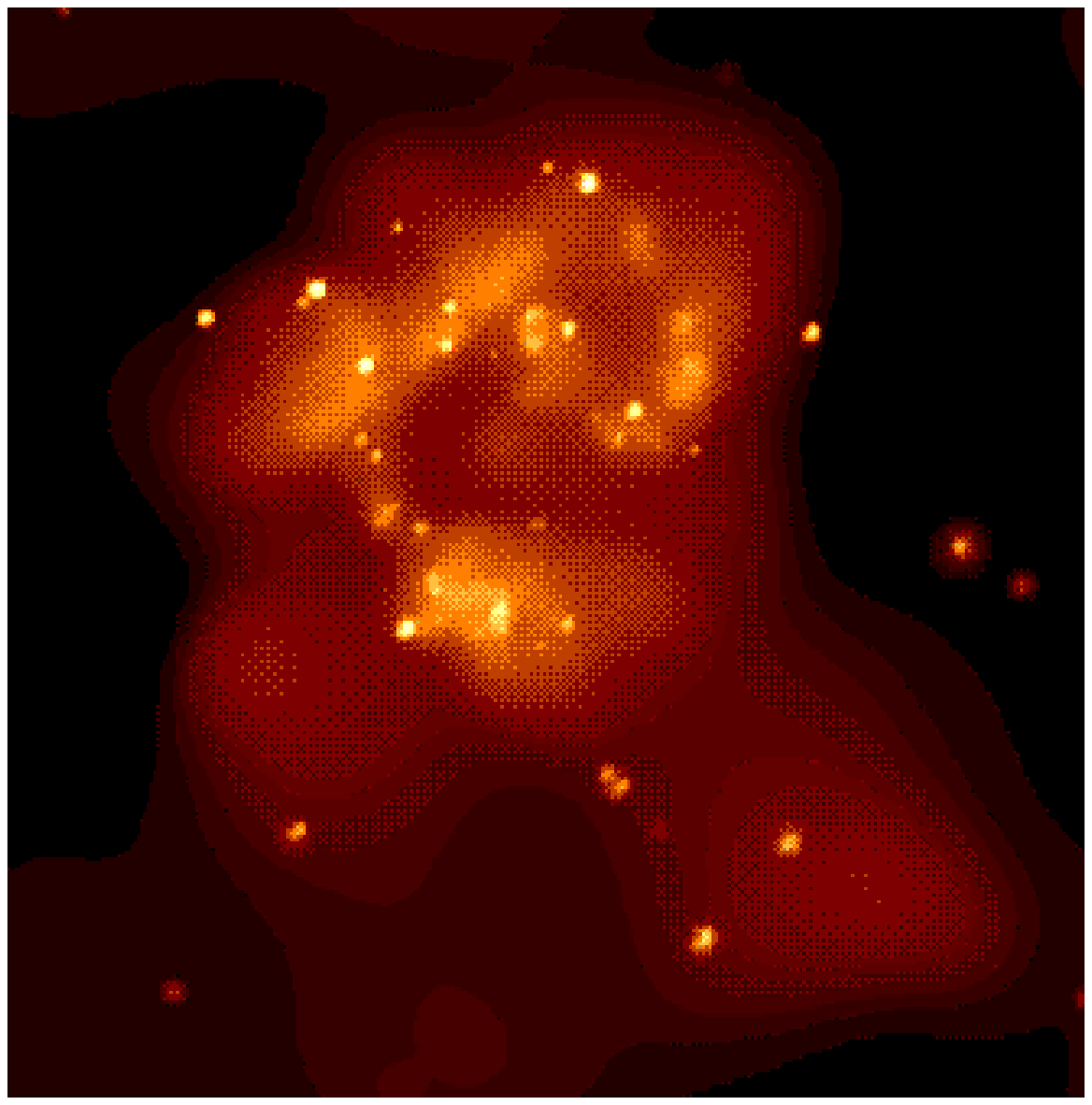,width=6.cm,angle=0.0} 
  \psfig{figure=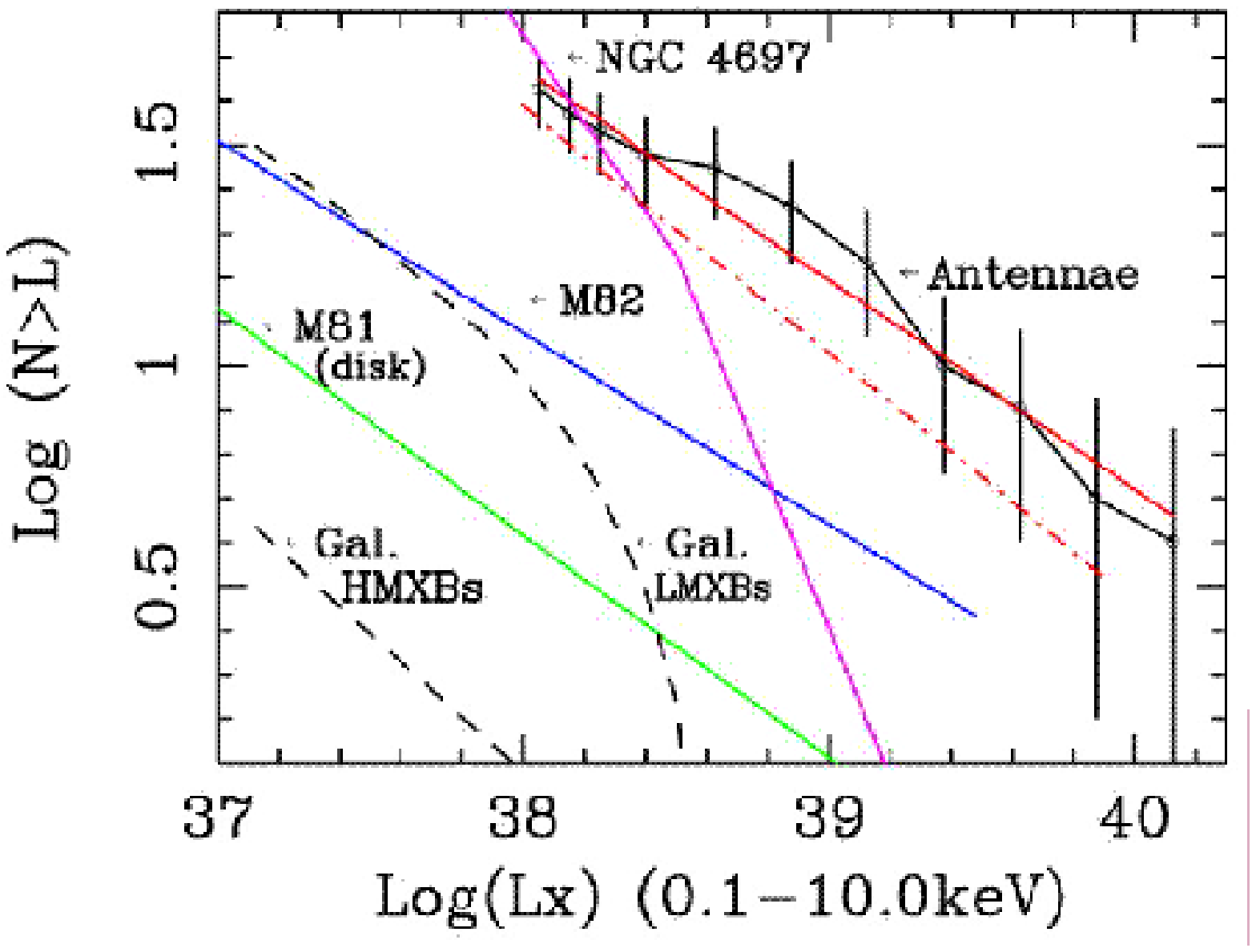,width=6.5cm,angle=0.0}
 }\par
  %,bbllx=1.5cm,bblly=13.7cm,bburx=15.5cm,bbury=25.7cm,clip=}}\par
  \caption{\label{fig:ant}
  Left: {\it Chandra} ACIS image of The Antennae (Fabbiano {\it et al.} 2001);
  Right: the XLF of The Antennae (points with error bars) compared with other
  galaxies, as labelled. Note the steep XLFs of the Galactic HMLXBs (bulge) 
  and of the early-type galaxy NGC~4697 (Zezas \& Fabbiano 2002).}
\end{figure*}

Observations show flatter XLF slopes (i.e., an increased presence of very luminous sources)
in galaxies with more intense star formation. The best example is given by the merger 
system NGC~4038/39 (The Antennae), where nine ultra-luminous X-ray 
sources (ULXs; $L_X > 10^{39} ~\rm ergs~s^{-1}$, for a distance of 19~Mpc)
were discovered with {\it Chandra} (Fabbiano, Zezas \& Murray 2001). 
Other examples of exceptionally luminous sources
are found in  M82 (Kaaret {\it et al.} 2001; Matsumoto {\it et al.} 2001), the Circinus galaxy 
(Smith \& Wilson 2001; Bauer {\it et al.} 2001) and NGC~1365~X-1 (Komossa \& 
Schultz 1998).
Consequently, flatter XLFs  occur in galaxies with more intense
star formation: the cumulative XLF slope is
- 0.45 in The Antennae
(Zezas \& Fabbiano 2002; Kilgard {\it et al.} 2002; Fig.~\ref{fig:ant}). 

Grimm,  Gilfanov \& Sunyaev (2003) suggest that the XLFs of star forming 
galaxies scale with the star formation rate (SFR), thus advocating
that HMXBs may be used as a star formation indicator in galaxies.
They find that at high SFRs the total X-ray luminosity
of a galaxy is linearly correlated to the SFR, and suggest a
`universal' XLF of starforming galaxies described by a power law with cumulative
slope of $\sim -0.6$ and a cut-off at $L_X \sim$few$\times 10^{40} ~\rm ergs~s^{-1}$.
This result of course depends on how well is the 
SFR of a given galaxy
known. This is a subject of considerable interest at this point, since various
indicators are differently affected by extinction.
The conclusion of a universal slope of the XLF of starforming galaxies
may be at odd with the reported correlation between the XLF slope and the 60$\mu$m 
luminosity from a minisurvey of spiral and starburst galaxies observed with 
{\it Chandra} (Kilgard {\it et al.} 2002).
 Also, theoretical models (Kalogera et al. 2003)
suggest that XLF slopes depend on the age of the starburst, so it is possible that
the `universal' XLF slope is not truly universal, but reflects a selection 
bias, in  that the sample used by Grimm, Gilfanov \& Sunyaev (2003)
may be dominated by starburtsts of similar ages.

\begin{figure*}[htbp]
  \vbox{\hskip -\leftskip
  \psfig{figure=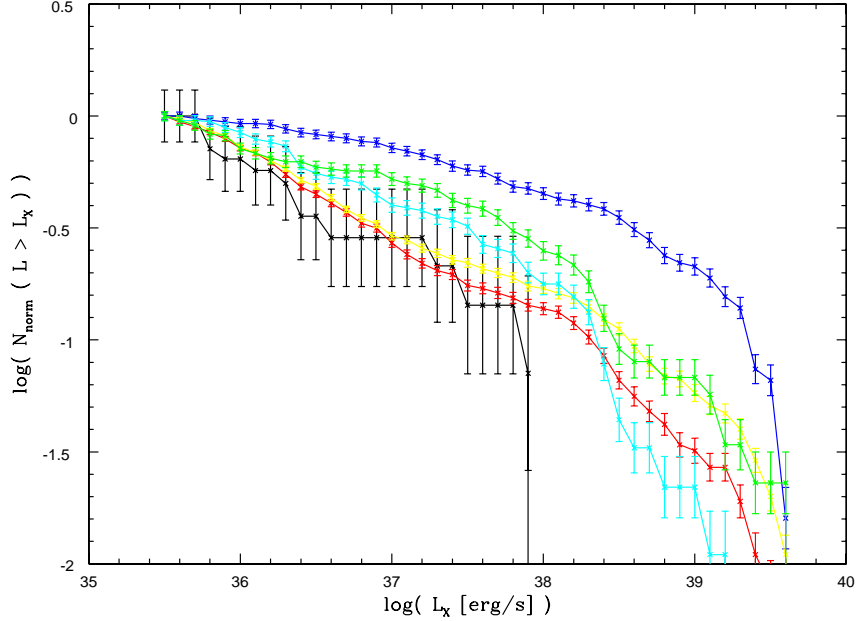,width=12.cm,angle=270.0} 
 }\par
  %,bbllx=1.5cm,bblly=13.7cm,bburx=15.5cm,bbury=25.7cm,clip=}}\par
  \caption{\label{fig:XLF}
  Comparison of XRB population models (from Kalogera et al. 2003) with the
observed XLF of NGC 1569 (bottom points, in black; data taken from Martin et al.\ 2002,
ApJ, 574, 663). Models were constructed to match the star-formation
history of NGC 1569 (recent star-burst duration and metallicity) and model
XLFs are shown at different times since the beginning of the starburst. Top to
bottom:
10Myr (blue), 50Myr (yellow), 110Myr (red), 150Myr (cyan), 200Myr (green).
Note that based on observations in other wavelengths, the age of the
starburst is estimated to be 105-110Myr.}
\end{figure*}

 Comparison of the XLFs for different 
galaxies, and modeling of the same, provide powerful tools for understanding 
the nature of the X-ray sources and for relating them to the evolution of the 
parent galaxy and its stellar population. Early theoretical work
has attempted to interpret the XLFs, using {\it ad hoc} power-law models, 
and accounting for aging and impulsive birth of XRB populations (Wu 2001, Kaaret 2002,
Kilgard {\it et al.} 2002).
Spurred by the recent observational developments, Kalogera and
collaborators have developed the first models of synthetic XLFs, based on
XRB evolutionary calculations (Belczynski et al. 2003). Such models
provide us with a potentially powerful tool for studying the origin and
evolution of XRB populations in stellar systems and their connection to
galactic environments. A preliminary examination of such models for
starburst galaxies (Kalogera et al. 2003; see Fig.~\ref{fig:XLF}) successfully shows that
predictions and consistency checks for the shapes and normalizations of
XLFs are possible with theoretical XRB modeling. These new developments
demonstrate that the predictions of 1995 are coming true (see Section~\ref{intro}).

\section{Ultra Luminous X-ray Sources - ULXs \label{ulx}}

ULXs are also named super-Eddington sources (see Fabbiano 1989, 1995), 
super-luminous sources, and intermediate luminosity X-ray objects (IXOs)  
(Roberts \& Warwick 2000; Colbert \& Mushotzky 1999;
Colbert \& Ptak 2003). All these names aim to convey the fact that
they  are extremely luminous X-ray sources, emitting
well in excess of the Eddington luminosity of a spherically  accreting and 
emitting neutron star ($\sim 2 \times 10^{38} ~\rm ergs~s^{-1}$).
Usually, sources emitting at $\sim 10^{39} ~\rm ergs~s^{-1}$
or above are included in this category. If these sources are emitting
isotropically at the Eddington limit, masses in excess of those expected 
from stellar black holes are implied, up to in some cases, $\geq 100 M_{\odot}$
(e.g. Fabbiano 1989, 1995; Makishima {\it et al.} 2000). Colbert \& Mushotzky (1999)
dubbed this type of black holes  `intermediate mass black holes' (IMBH), to distinguish 
them from the stellar mass black holes found in Galactic black hole binaries,
and also from the supermassive $10^7-10^9 M_{\odot}$ found at the nuclei
of galaxies that are responsible for AGNs.

\subsection{Spectra and spectral variability}

\begin{figure*}[htbp]
  \vbox{\hskip -\leftskip
  \psfig{figure=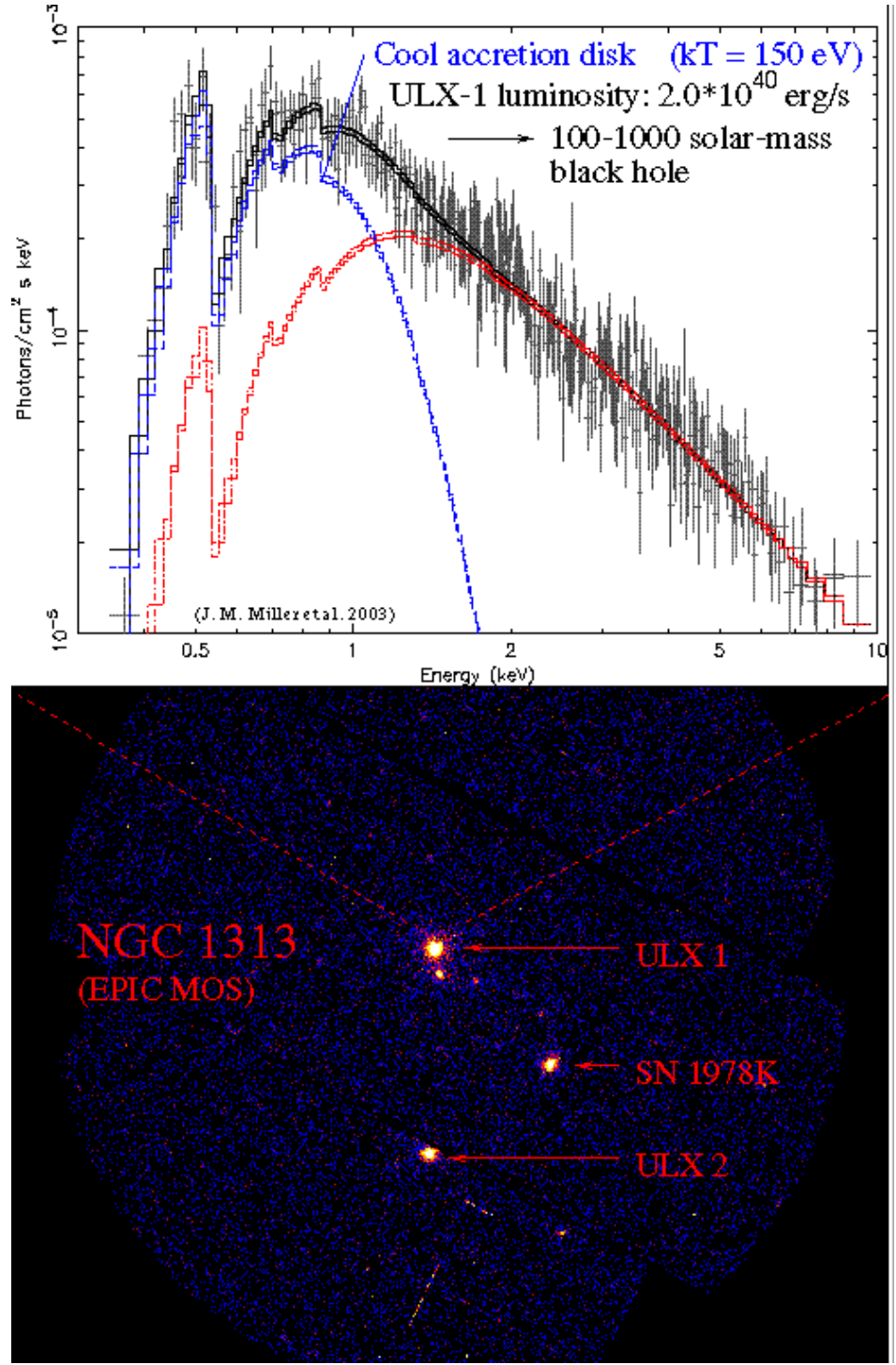,width=12.5cm,angle=0.0}
  }\par
  %,bbllx=1.5cm,bblly=13.7cm,bburx=15.5cm,bbury=25.7cm,clip=}}\par
  \caption{\label{fig:miller}
 Bottom: the {\it XMM-Newton} image of NGC~1313, showing the position of the 
 two ULXs. Top: X-ray spectrum of ULX-1, compared with best-fit model requiring
 a cool accretion disk component (Miller {\it et al.} 2003).}
\end{figure*}

Although young supernova remnants may be responsible for ULX emission
in some cases (e.g Fabian \& Terlevich 1996), there is now sufficient evidence 
from spectral and variability data, to establish that
the majority ULXs are indeed compact systems, most likely accreting binaries.
 {\it ASCA} X-ray spectra suggested accretion disk 
emission. These spectra, however, also require
temperatures much larger than those expected from black holes of the mass
implied by the luminosities of these sources, leading to the suggestion 
of rotating Kerr black holes (Makishima {\it et al.} 2000; Mizuno, Kubota \&
Makishima 2001). 
In The Antennae ULXs the {\it Chandra} spectra (Zezas {\it et al.} 2002a, b) tend to be hard, and
their average co-added spectrum requires both a power law ($\Gamma \sim 1.2$) and a  
disk-blackbody component consistent with the {\it ASCA} results, with kT$\sim 1.1$~keV.
A {\it {\it XMM-Newton}} survey of 10 galaxies reports ULX spectra consistent
with black hole binaries in either high or low state (Foschini {\it et al.} 2002), but
the data quality is too poor for detailed modelling.
Similar general  spectral results can be found in a {\it Chandra} survey of ULXs in
different galaxies (Humphrey {\it et al.} 2003). 
Instead, {\it {\it XMM-Newton}} high quality spectra of two ULXs in NGC~1313 (X-1 and X-2) led to highly
significant detections of soft accretion disk components, with temperatures of kT$\sim 150$~eV,
consistent with accretion disks of IMBHs (Miller {\it et al.} 2003a; Fig.~\ref{fig:miller}).

\begin{figure*}[htbp]
  \vbox{\hskip -\leftskip
  \psfig{figure=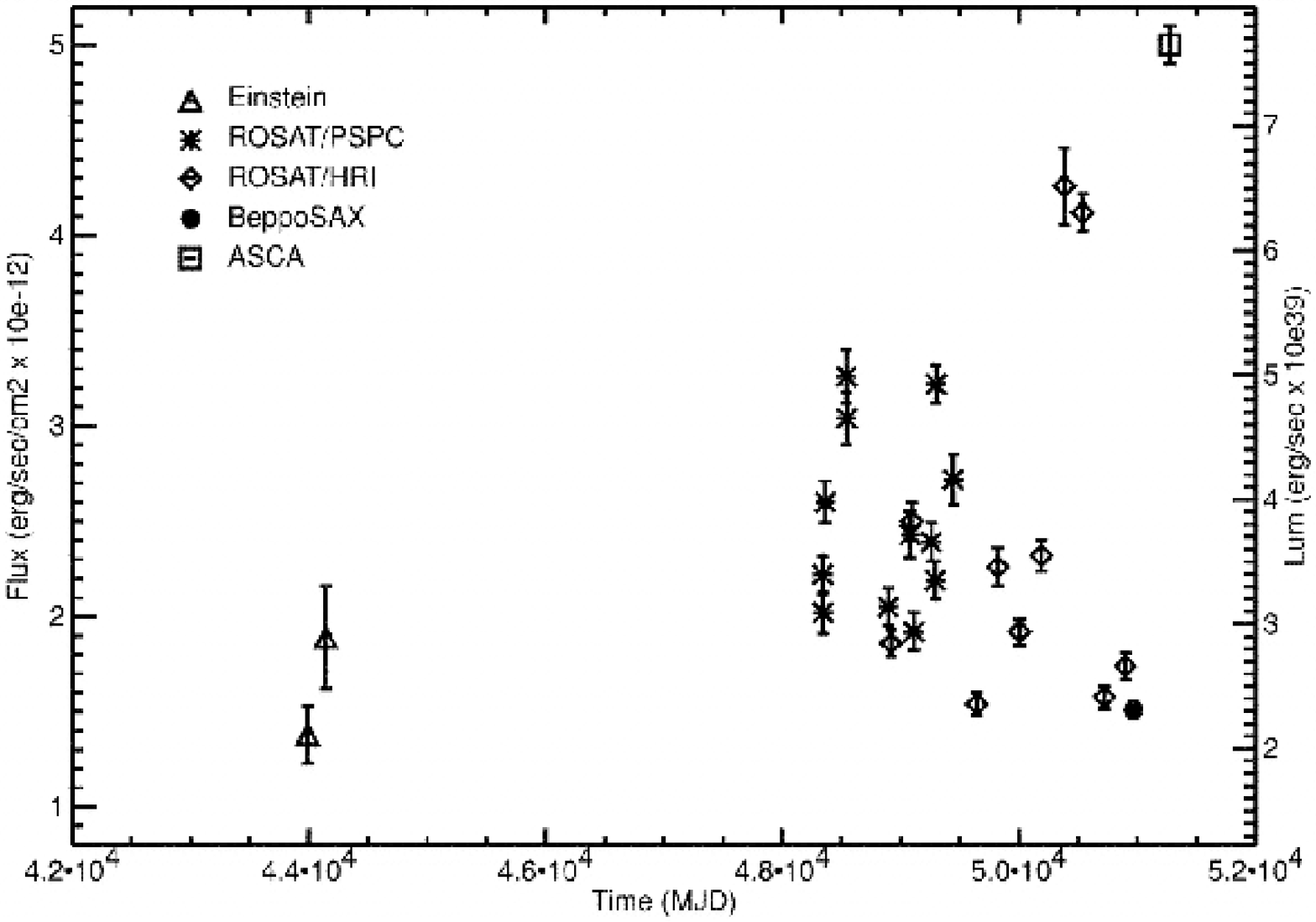,width=12.5cm,angle=0.0}
  }\par
  %,bbllx=1.5cm,bblly=13.7cm,bburx=15.5cm,bbury=25.7cm,clip=}}\par
  \caption{\label{fig:x9}
  Light-curve of M81~X-9, covering $\sim 20$~yrs of observations (La~Parola {\it et al.} 2001).}
\end{figure*}

The XRB hypothesis
is reinforced by observations of correlated luminosity-spectral
variability similar to the  `high/soft-low/hard' behavior of Cyg~X-1
(e.g., in M81~X-9, La~Parola {\it et al.} 2001, with a variety of X-ray 
telescopes, Fig.~\ref{fig:x9}; and in two ULXs in IC~342,
Kubota {\it et al.} 2001 with {\it ASCA}). 
However, more recently, Kubota, Done \& Makishima (2002) argue that
these power-law ULX spectra should not be identified with the low/hard state,
but rather may be due to a strongly Comptonized optically thick accretion disk,
analogous to the Comptonization-dominated `very high/anomalous state' in
Galactic black-hole binaries.
{\it ASCA} observations of one of the IC~342 sources in high
state (disk-dominated) revealed a `high/hard-low/soft' low-level
variability, with a possible 30-40~hr periodicity, as could be produced
by a massive main sequence star orbiting a black hole
(Sugiho {\it et al.} 2001). 

With {\it Chandra} and {\it XMM-Newton}  an increasing number of ULXs
are being discovered and studied in galaxies. Variability in the {\it Chandra}
observations of M82 established that the ULXs in this galaxy are
likely to be accreting compact objects (Matsumoto {\it et al.} 2001).
The {\it Chandra} observations of NGC~3628 (Strickland {\it et al.} 2001) show the re-appearance 
of the $10^{40} ~\rm ergs~s^{-1}$ variable ULX first discovered with {\it {\it ROSAT}}
(Dahlem, Heckman \& Fabbiano 1995). A new transient ULX was discovered in M74 (NGC~628) with 
(Soria 
\& Kong 2002). {\it Chandra} observations of MF~16 in NGC~6946, formerly 
identified as an extremely luminous supernova remnant (Schlegel 1994), 
reveal instead a point-like source with the typical X-ray spectrum of a 
black-hole binary (Holt {\it et al.} 2003; Roberts \& Colbert 2003).
Similarly, M81~X-6, which is positionally coincident with a supernova
remnant, is identified as a XRB by its X-ray spectrum (Swartz {\it et al.} 2003).
{\it Chandra} observations
of the nucleus of M33 have revealed a two-component (power-law and disk)
spectrum and have established luminosity-spectral 
variability patterns in this ULX, reminiscent of the black hole 
binary LMC X-3 (La Parola {\it et al.} 2003; see also Long, Charles \& Dubus 2002);
Dubus \& Rutledge (2002) compare this source with the Galactic 
microquasar GRS~1915+105.
 
\begin{figure*}[htbp]
  \vbox{\hskip -\leftskip
  \psfig{figure=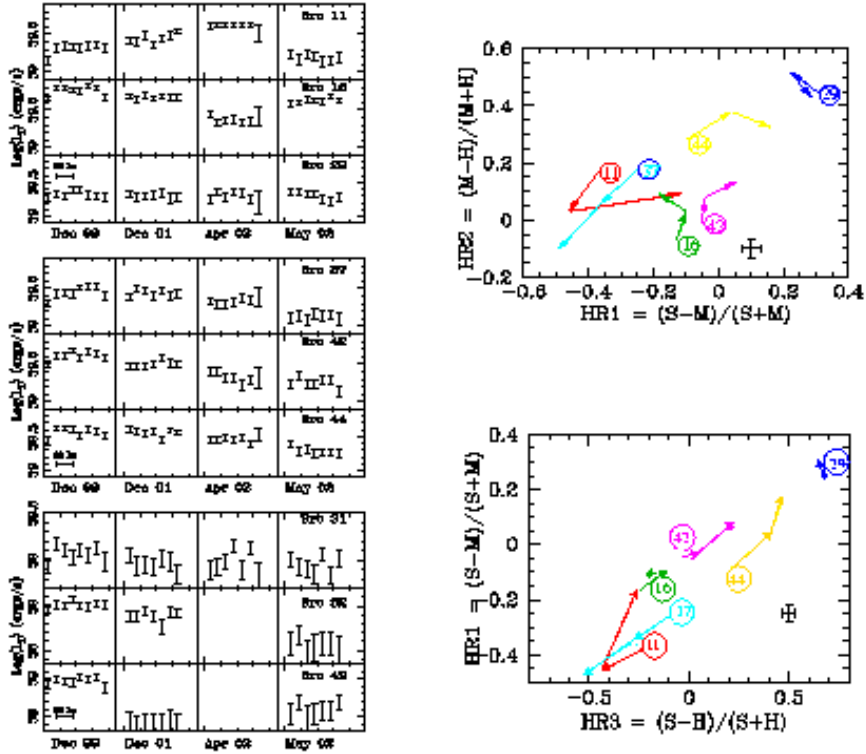,width=12.5cm,angle=0.0}
  }\par
  %,bbllx=1.5cm,bblly=13.7cm,bburx=15.5cm,bbury=25.7cm,clip=}}\par
  \caption{\label{fig:antvar}
  Left: {\it Chandra} light curves of the ULXs of The Antennae. Right:
  color-color diagrams of the most luminous sources (Fabbiano {\it et al.} 2003a).}
\end{figure*}

High/hard-low/soft variability
was found in M51~X-7, together with a possible 2.1~hr period (but the
time coverage is scant) by Liu {\it et al.} (2002). Both Cyg~X-1 like  high/soft-low/hard
as well as  high/hard-low/soft variability was detected in the population
of nine ULXs discovered with {\it {\it Chandra} } in the Antennae galaxies
(Fabbiano ; Fig.~\ref{fig:antvar}). The latter type of variability can also be found
in a few Galactic XRBs (1E~1740.7-2942, GRS~1758-258, GX~339-4, Smith {\it et al.}\ 2002; 
see also the {\it XMM-Newton} results on GRS~1758-258, Miller {\it et al.}\ 2002).
This spectral variability may be indicative of the competition
between the relative dominance of the accretion disk versus the innermost
hot accretion flow; several scenarios for spectral variability are
discussed in Fabbiano {\it et al.} 2003a  and references therein.

\subsection{Intermediate Mass Black Holes or Beamed XRBs?}

Although there is clear evidence pointing towards an XRB nature for ULXs,
the presence of IMBHs in these systems is by no means universally accepted,
and it may be quite possible that ULXs are indeed a heterogeneous population.
As discussed above, the {\it ASCA} spectra were interpreted by Makishima {\it et al.} (2000)
as evidence for rotating Kerr IMBH, to reconcile the high accretion disk temperature
suggested by the model fitting of these spectra with the large black hole masses
implied by the bolometric luminosity of the ULXs, which  would require much cooler
disks for a non-rotating IMBH. Colbert \& Mushotzky (1999) suggested that these
cooler accretion disk components may be present in their {\it ASCA}
survey of ULXs, but the statistical significance
of these early claims is not very high. The {\it Chandra } detections of  
super-soft ULXs (e.g., Swartz {\it et al.} 2002, in M81; Di~Stefano {\it et al. 2003}
in M104; see also later in this Section) 
could be interpreted as evidence for IMBHs. More 
important, low-temperature components vere discovered in the
{\it {\it XMM-Newton}} spectra of `normal' ULXs: in the NGC~1313 ULXs, which do not require a 
Kerr black hole, and are  entirely consistent with emission from an 
IMBH accretion disk (Miller {\it et al.} 2003a; Fig.~\ref{fig:miller});  and 
in at least one of the ULXs in the Antennae galaxies (kT$\sim$0.13~keV)
(Miller {\it et al.} 2003b). 

Considerable attention has 
been devoted to an extremely luminous variable 10$^{40} ~\rm ergs~s^{-1}$ 
ULX detected with {\it Chandra} near 
the dynamical center of M82. In the picture of spherical accretion onto an
IMBH, the luminosity of this source would imply masses in excess of 100~$M_{\odot}$ for the accretor.
This ULX appears to be at the center of an expanding molecular superbubble
with 200~pc diameter (Matsushita {\it et al.} 2000).
Based on  its accurate {\it Chandra} position, 
which  is not at the nucleus, Kaaret {\it et al.} (2001) set an upper limit of $10^5-
10^6 M_{\odot }$ to its mass. Strohmayer \& Mushotzky (2003)
report quasi
periodic oscillations (QPOs) in the {\it XMM-Newton} data of this source. They argue that 
their discovery suggests emission from an accretion disk and is incompatible with 
the radiation being beamed, and therefore implying a less extreme 
emitted luminosity, as in King {\it et al.} (2001; see below). On the assumption that the highest QPO
frequency is associated with the Kepler frequency at the innermost circular orbit around a Schwarzschild 
black hole, these authors set an upper limit of 1.87$ \times 10^4 M_{\odot }$ to 
the black hole mass: this source could therefore be an IMBH, with masses in
the 100-10,100~$M_{\odot}$ range. However, as noted by Strohmayer \& Mushotzky (2003),
the crowded M82 field cannot be spatially resolved with {\it XMM-Newton}, 
making the association of the QPO with the most luminous ULX in the field not
entirely proven. Moreover, the spectral fit of these data suggests a temperature 
kT$\sim 3$~keV, much higher than the one expected from an IMBH accretion disk.

As we will discuss below, some results are hard to explain in the IMBH scenario. 
Two other models have been advanced,  which do
not require IMBH masses. The large number of ULXs found in The Antennae
led to the suggestion that they may represent a normal stage of XRB evolution (King {\it et al.}
2001). In the King {\it et al.} (2001) model, the apparent (spherical) accretion luminosity is boosted
because of geometrical collimation of the emitting area in thick accretion disks,
resulting from the large thermal-timescale mass transfer characterizing the later 
stages of a massive XRB (see Chapter by King in this book). Exploiting the similarity with Galactic microquasars,
the jet emission model of K\"{o}rding {\it et al.} (2002) produces enhanced luminosity via
relativistic beaming. In at least one case, the variable luminous ULX 2E1400.2-4108 in
NGC~5408, there is observational evidence pointing to this relativistic jet model:
Kaaret {\it et al.} (2003) find weak radio emission associated with the X-ray source, and argue 
that the both the multi-wavelength spectral energy distribution, and the X-ray spectrum
are consistent with the K\"{o}rding {\it et al.} (2002) scenario.

In some cases at least the IMBH hypothesis is supported by the association of the ULX with
diffuse H$\alpha$ nebulae, suggesting isotropic illumination of the interstellar
medium by the ULX, and therefore absence of beaming (e.g. Pakull \& Mirioni 2002
in the case of the NGC~1313 sources, see Miller {\it et al.} 2003).
M81~X-9 is also associated with an optical nebula, which also contains
hot gas (La~Parola {\it et al.} 2001; Wang 2002). Wang (2002) considers the possibility 
that this nebula may be powered by the ULX and also speculates that it
may be the remnant of the formation of the ULX.
Weaver {\it et al.} (2002) discuss a heavily absorbed ULX in the nuclear starburst of 
NGC~253; this source appears to photoionize the surrounding gas. Weaver {\it et al.}
speculate that it may be an IMBH, perhaps connected with either the beginning
or the end of AGN activity. However, in at least one case (IC~342~X-1, Roberts {\it et al.} 2003),
there is a suggestion of anisotropic photoionization, 
that may indicate beamed emission from the ULX.

In The Antennae, comparison with $\it {\it HST}$ data shows that 
the ULXs are offset from  starforming
stellar clusters. While coincidence with a stellar cluster may be due to happenstance
because of the crowded fields, the absence of an optical counterpart is a solid result and suggests
that the ULXs may have received kicks at their formation (Zezas \& Fabbiano 2002), 
which would be highly unlikely
in the case of a massive IMBH forming in a dense stallar cluster (e.g. Miller \& Hamilton 2002).
An alternate IMBH scenario, discussed by Zezas \& Fabbiano, is that of
primordial IMBHs drifting through stellar clusters after
capturing a companion (Madau \& Rees 2001).

Other optical studies find counterparts to ULXs, and set indirect constraints 
on the nature of the accretor.
A blue optical continuum counterpart to the variable ULX NGC~5204~X-1 was
found by Roberts {\it et al.} (2001), and subsequently resolved by Goad {\it et al.} (2002) with
{\it {\it HST}}. These authors conclude that the stellar counterpart points to an
early-type binary. Similarly, Liu, Bregman \& Seizer (2002) find an 08V star conterpart for 
M81~X-1, a ULX with average $L_X \sim 2 \times 10^{39} ~\rm ergs~s^{-1}$.
These counterparts may be consistent with the picture of King {\it et al.} (2001), of ULXs 
as XRBs experiencing thermal timescale mass transfer.

Recent results on supersoft variable ULXs suggest that the emitting region may not be
associated with the inner regions of IMBH accretion disks in these sources, but may be
due to Eddington-driven outflows from
a stellar mass black hole. The spectral variability (at constant bolometric luminosity) of the soft ULX P098
in M101 (detected with {\it Chandra}; Mukai {\it et al.} 2003) led to the suggestion of 
an optically thick outflow 
from a 15-25~\msun\ black hole, regulated by the Eddington limit.
{\it Chandra} time monitoring observations of The Antennae have led to the discovery
of  a  variable super-soft source (kT = 90 - 100~eV for
a blackbody spectrum), reaching ULX luminosities of 2.4$\times 10^{40}
~\rm ergs~s^{-1}$ (Fabbiano
{\it et al.} 2003b). The assumption of unbeamed
emission would suggest a black hole of $\geq 100 M_{\odot}$. However the
radiating area would have to vary by a factor $\sim 1000$ in this
case, inconsistent with gravitational energy release from within a few
Schwarzschild radii of a black hole. 
As discussed in  (Fabbiano
{\it et al.} 2003b), a surprising possible solution is a white dwarf with $M \sim 1$\msun, at the
Eddington limit, with a variable beaming factor (up to a beaming factor $b \sim 10^{-2}$). 
A second possible solution involves outflows from a stellar--mass black hole,
accreting near the Eddington limit (as in Mukai
{\it et al.} 2002) but with mildly anisotropic radiation
patterns ($b \sim 0.1$, as in King {\it et al.} 2001). Similar sources are
reported in  M81 (Swartz {\it et al.} 2002), NGC~300 (Kong \& Di~Stefano 2003),
and other nearby spiral galaxies (Di~Stefano \& Kong 2003; see also Di~Stefano {\it et al}
(2003) for SSSs in M31).

Transient behavior has been shown to be an important observational
diagnostic that could allow us to distinguish between beamed models and IMBH
accretion for the origin of ULXs in young, star-forming regions (Henninger
et al. 2003). Accretion onto IMBH black holes can lead to unstable disks
and hence transient behavior whereas beamed binary systems have transfer
rates that are high enough for the disks to be stable and X-ray emission
to be persistent. Therefore long-term monitoring can prove a
valuable and possibly unique tool in unraveling the nature of ULXs.

\section{XRBs in Elliptical and S0 Galaxies \label{ell}} 

As discussed in the 1995 chapter (Fabbiano 1995), XRBs could
not be directly detected in E and S0 galaxies with pre-{\it Chandra}
telescopes, because of the distance of these galaxies and the
limited angular resolution of the telescopes.
The presence of XRBs in E and S0 galaxies was predicted by Trinchieri \&
Fabbiano (1985), based on an analogy with the bulge of M31, for which such a 
population could be detected (Van Speybroeck {\it et al.} 1979; see also
Fabbiano, Trinchieri \&
Van Speybroeck 1987). This early claim was
reinforced by differences in the average spectral properties
of E and S0 galaxies with different X-ray-to-optical luminosity 
ratios, that suggested a baseline X-ray faint XRB emission
(Kim, Fabbiano \& Trinchieri 1992; Fabbiano, Kim \& Trinchieri 1994), 
and by the {\it ASCA} discovery of a hard spectral 
component in virtually all E and S0 galaxies (Matsushita {\it et al.} 1994),
which, however, could also have been due, at least in part, to accreting 
massive nuclear black holes (Allen, Di Matteo \& Fabian, 2000).

\begin{figure*}[htbp]
  \vbox{\hskip -\leftskip
  \psfig{figure=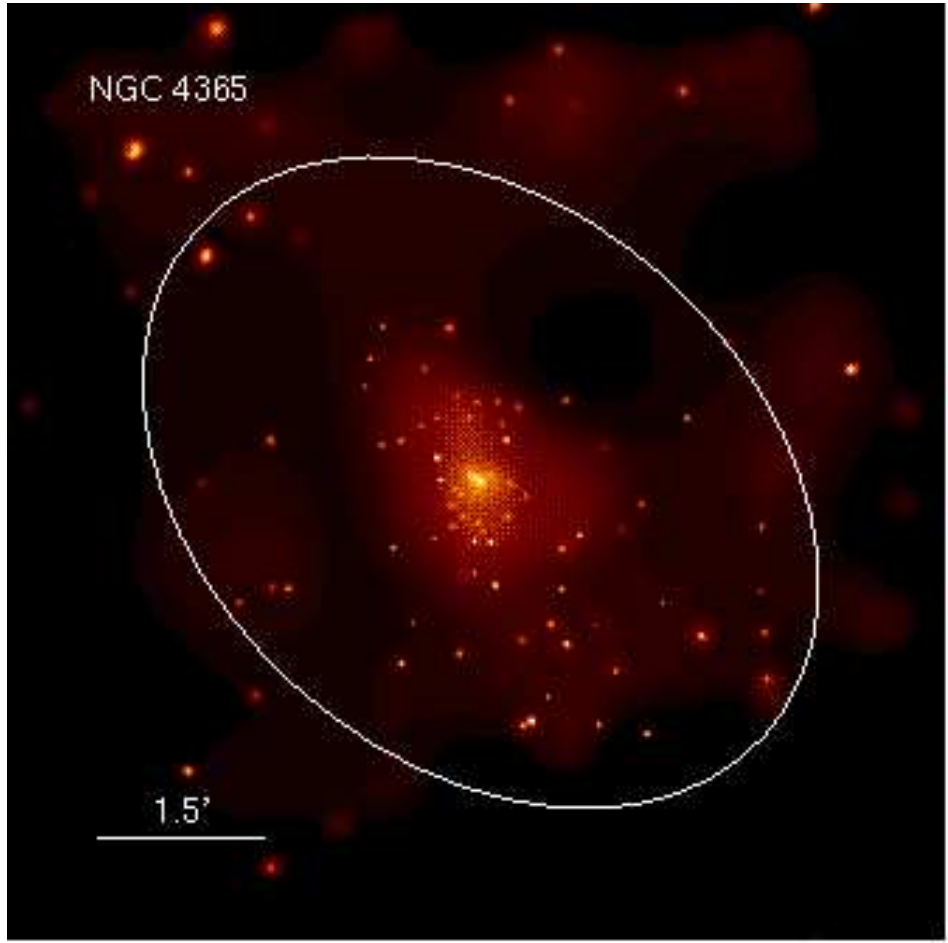,width=12.5cm,angle=0.0}
  }\par
  %,bbllx=1.5cm,bblly=13.7cm,bburx=15.5cm,bbury=25.7cm,clip=}}\par
  \caption{\label{fig:4365}
  {\it Chandra} ACIS image of the Virgo elliptical NGC~4365, using archival data.
  The white ellipse is the $D_{25}$ isophote, from de~Vaucouleurs {\it et al.} (1991).}
\end{figure*}

The {\it Chandra} images (Fig.~\ref{fig:4365}) leave no doubt about the presence of rich populations
of point-like sources in E and S0 galaxies. Published results, of which the first one
is the paper on NGC~4697 by Sarazin, Irwin \& Bregman (2000),
include point-source detections in a number of galaxies. These
source populations have been detected with 
varying low-luminosity detection thresholds (a function of galaxy distance and
observing time). While most of the detected sources have luminosities in the
$10^{37} - 10^{39} ~\rm ergs~s^{-1}$ range,  some were detected at luminosities above 
$10^{39} ~\rm ergs~s^{-1}$, in the Ultra-Luminous-X-ray (ULX) source range
(see Section~\ref{ulx}). A
representative summary (limited to papers published or in press as of May 2003) in given in Table~11.1.

\small
\begin{table}
\caption{E \& S0 Galaxies: Representative Summary of {\it Chandra} Results }
\begin{tabular}{lccl}
\hline
Name & No. of  & $L_X~(\rm ergs~s^{-1})$ & Comment\\
&sources& band (keV)& \\
\hline
NGC~720&42 & $4\times 10^{38} - 1\times 10^{40}$ &
9 ULX in `arc' pattern\\
&&0.3-7& 12
associations with GCs\\
&&& (Jeltema {\it et al.} 2003)\\\\
NGC~1291 & $\sim$50 & $< 3 \times 10^{38}$ & 3 associations with GCs \\
&&0.3-10&(Irwin {\it et al.}. 2002)\\\\
NGC~1316 & 81 & $2\times 10^{37} - 2\times 10^{39}$& kT$\sim$5~keV average spectrum\\
&&0.3-8&5 associations with GCs\\
&&&(Kim \& Fabbiano 2003)\\\\
NGC~1399 & $\sim 140$ & $5\times 10^{37} - 5\times 10^{39} $& 70\% associated with GCs \\
&&0.3-10&(Angelini {\it et al.}. 2001)\\\\
NGC~1553& 49 & $ 1.6\times 10^{38} - \sim\times 10^{40}$ &X-ray colors consistent \\
&&0.3-10&with NGC~4697\\
&&&3 associations with GCs\\
&&&(Blanton {\it et al.}. 2001)\\\\
NGC~4374& $\sim 100$ & $ 3\times 10^{37} - \sim 2\times 10^{39} $ & spectra consistent \\
(M84)&&0.4-10&with Galactic LMXB\\
&&&(Finoguenov \& Jones 2002)\\\\
NGC~4472& $\sim 120$ & $ 1\times 10^{37} - \sim 1.5\times 10^{39} $ & 40\% associated with CGs\\
&&0.5-8&(Kundu {\it et al.}. 2002)\\\\
NGC~4697& $\sim$80 & $5\times 10^{37} - 2.5\times 10^{39} $ & average spectrum kT$\sim$8~keV\\
&&0.3-10&7 (20\%) in GCs \\
&&&(Sarazin {\it et al.}. 2001)\\\\
NGC~5128 & 246 & $2\times 10^{36} - 1\times 10^{39}$& 9 identifications with GCs \\
(CenA)&&0.4-10&(Kraft {\it et al.} 2001)\\\\
NGC~5846 & $\sim 40$ &$3\times 10^{38} - 2\times 10^{39} $& (Trinchieri \& Goudfrooij 2002)\\
&&0.3-10&\\
\hline
 \end{tabular}
 \end{table}
\normalsize

 \medskip
 The X-ray colors or co-added spectra of these sources  are consistent with those of
 LMXBs (see above references, and Irwin, Athey \& Bregman 2003); 
 however, a variety of spectral properties have been reported in some cases,
 similar to the spectral variety of Galactic and Local Group XRBs, including a few instances of
 very soft and supersoft (i. e., all photons below $\sim 1~$keV) sources (e.g., NGC~4697, 
 Sarazin, Irwin \& Bregman 2000; M84, Finoguenov \& Jones 2002; NGC~1399, Angelini,
 Loewenstein \& Mushotzky 2001; NGC~1316, Kim \& Fabbiano 2002). The overall spatial distribution
 of these sources follows that of the stellar light, but there are
 exceptions, such as in NGC~720, where the most luminous sources follow arcs 
 (Jeltema {\it et al.} 2003), NGC~4261 and NGC~4697, where the X-ray source distributions
 are highly asymmetric (Zezas {\it et al.} 2003), 
 and NGC~4472, where the X-ray source distribution may be more consistent with
 that of Globular Clusters (GCs) than of the general field stellar light (Kundu, Maccarone
 \& Zepf 2002;
 Maccarone, Kundu \& Zepf 2003). No firm conclusion on the
 origin and evolution of these sources exists. Given the old stellar population of the parent
 galaxies, and the life-times of LMXBs, it has ben suggested that these sources may be 
 outbursting transients (Piro \& Bildsten 2002). Alternatively, more recent formation and evolutions
 in GCs may result in steady sources (Maccarone, Kundu \& Zepf  2003). 
 With the exception of NGC~5128, which is near enough to allow detection of
sources in the $10^{36} ~\rm ergs~s^{-1}$ luminosity range, and for which multiple observations
demonstrate widespread source variability (Kraft {\it et al.} 2001), 
the {\it Chandra} observations performed so far typically 
only give a single snapshot of the most luminous part of the XRB population in a given galaxy.
In NGC~5128, a comparison of the two {\it Chandra} observations reveals at least five transients 
(sources that disappear with a dimming factor of at least 10), supporting the Piro \& Bildsten
scenario.

{\it Chandra} observations of highly significant asymmetries in the spatial distribution of
X-ray sources in otherwise regular old elliptical galaxies (Zezas {\it et al.} 2003) may 
suggest rejuvenation of the stellar population of these galaxies. In NGC~4261, the most 
significant example, all the detected sources are luminous, above the Eddington limit for a
neutron star accretor. If the X-ray sources were standard LMXBs belonging to the dominant
old stellar population, we would expect their spatial distribution to be consistent (within
statistics) with that of the stellar light. However this is not so, as indicated by Kolmogorov-Smirnov
tests and Bayesian block analysis. On the basis  of simulations of galaxy interactions
(Hernquist \& Spergel 1992; Mihos \& Hernquist 1996),
this result suggests that the luminous XRBs may belong
to a younger stellar component, related to the rejuvenating fall-back of material in tidal
tails onto a relaxed merger remnants.
 
 \subsection{ULXs in Early-Type Galaxies }

As can be seen from Table~11.1, in early-type galaxies the 
occurrence of sources with $L_X = 1-2 \times
10^{39} ~\rm ergs~s^{-1}$  is common, 
although generally limited to a few sources per galaxy. 
These sources could easily be explained with normal black hole binaries or
moderately beamed neutron star binaries (King 2002).
In their mini-survey of 14 galaxies observed with {\it Chandra} 
(which include some of the ones listed in Table~11.1), Irwin, Athey \& Bregman (2003) 
find that of the four sources with X-ray luminosities
in the $1-2 \times 10^{39} ~\rm ergs~s^{-1}$ range for which they can derive spectra,
three have soft spectra, similar to those of black hole binaries in high 
state (see also Finoguenov \& Jones 2002). 

Not much can be said about the variability of ULXs in early-type galaxies,
because repeated {\it {\it Chandra} } observations of a given galaxy are not
generally available.  
In the case of NGC~5128, comparison with previous {\it {\it ROSAT}} images (see Colbert and Ptak 2002)
shows considerable flux variability in these very luminous sources: two ULXs were
detected in {\it {\it ROSAT}} observations, both have
considerable lower luminosities in the {\it Chandra} data (Kraft {\it et al.} 2001), and one of them
may have disappeared.

While, in general, sources with $L_X > 2 \times 10^{39} ~\rm ergs~s^{-1}$ are relatively
rare in early-type galaxies as compared to actively star-forming galaxies (see Section~\ref{ulx}),
and may be preferentially
associated with GCs (e.g. Angelini, Loewenstein \& Mushotzky 2001; see 
Irwin, Athey \& Bregman 2003), this is not always the case, as exemplified by NGC~720.
This galaxy  (Jeltema {\it et al.} 2003) is peculiar in possessing nine ULXs
(this number is of course dependendent on the assumed distance, 35~Mpc), a population as rich as
that of the actively starforming merger galaxies The Antennae 
(Fabbiano, Zezas \& Murray 2001, Zezas \& Fabbiano 
2002). 
Only three of these ULXs can be associated with GCs. The sources in 
NGC~720 are also peculiar in their spatial distribution, which does not follow the distribution 
of the optical light, as it would be expected from LMXBs evolving from low-mass bulge
binaries: these sources are distributed in arcs. Their large number and their spatial 
distribution may suggest that they are younger systems, perhaps the remnants of a recent 
merger event. 

The associations of some ULXs in early-type galaxies with GCs
may support the possibility that a subset of these sources may be associated with 
IMBH ($> 10 M_{\odot}$) (see Fabbiano 1989
and refs. therein; Irwin, Athey \& Bregman 2003). 
However, most of the ULXs in early-type galaxies
are likely to be lower mass binaries, given the stellar population of the parent galaxy.
King (2002; see also Piro \& Bildsten 2002) 
suggests that they may be a class of ULXs associated with outbursts of
soft X-ray transients, resulting in moderately beamed emission from the inner regions of a
thick accretion disk. In the case of NGC~720 they may be related to a `hidden' younger
stellar population (Jeltema {\it et al.} 2003).

 \subsection{X-ray sources and Globular Clusters  }
 
 The association of X-ray sources in early type-galaxies with GCs has been 
 widely discussed. As can be seen from Table~11.1, associations with GCs range from $\leq
 10\%$ in most galaxies, $\sim 40\%$ in some Virgo galaxies (NGC~4472, NGC~4649), to 70\% in
 NGC~1399, the dominant galaxy in a group . The statistics are somewhat fraught with uncertainty, 
 since lists of GCs from {\it {\it HST}} are not available  for all the galaxies
 studied with {\it Chandra}, and the detection thresholds differ in different galaxies.
  However, this association is interesting and has led to the suggestion that perhaps 
 all the LMXBs in early-type galaxies may form in GCs, from whence they 
 may be expelled if they receive strong enough kicks at their formation,
 or may be left behind if the GC is tidally disrupted. 
 This suggestion was first advanced by 
 Sarazin, Irwin \& Bregman (2000), and was more recently elaborated by 
 White, Sarazin \& Kulkarni (2002), on the basis of a correlation of the specific 
 GC frequency  with the ratio of the integrated LMXB luminosity
 to the optical luminosity of eleven galaxies. 
Kundu, Maccarone \& Zepf (2002) explore the LMXB-GC connection in NGC~4472, where
they find that  40\% of the sources detected at $L_X > 1 \times 10^{37} ~\rm ergs~s^{-1}$
are associated with GCs. In this galaxy, the fraction of GCs hosting an X-ray source is
 4\%, the same as in the Galaxy and M31. More luminous, more metal rich, and more centrally
 located GCs are more likely to host 
 LMXBs, reflecting both an increased probability of binary formation with the numbers
 of stars in a GC, and also an effect of metallicity in aiding binary formation
 (Kundu, Maccarone \& Zepf 2002).

 \begin{figure*}[htbp]
  \vbox{\hskip -\leftskip
  \psfig{figure=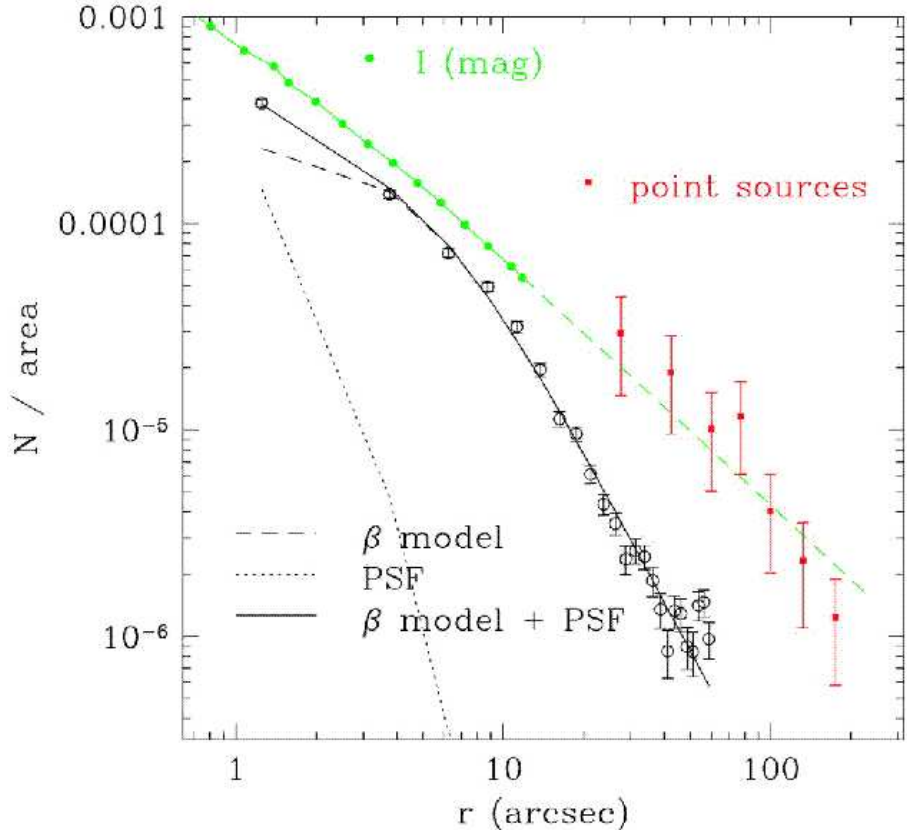,width=12.cm,angle=0.0}
  }\par
  %,bbllx=1.5cm,bblly=13.7cm,bburx=15.5cm,bbury=25.7cm,clip=}}\par
  \caption{\label{fig:1316distr}
  Radial distributions of the emission components of NGC~1316. The gaseous 
  component (hot ISM) is represented by the inner distribution of points.
  The cumulative XRB contribution is given by the outer set of points; the 
  dashed line through these points is the extrapolation of the stellar (I)
  surface brightness (Kim \& Fabbiano 2003).}
\end{figure*}

 While the possibility of LMXB formation in GCs is intriguing,
 this is still an open question, since evolution of bulge stars may also produce LMXBs
 (e.g., Kalogera \& Webbink 1998; Kalogera 1998). 
 The spatial distribution of the LMXBs, if it follows the optical stellar
 light (e.g. in NGC~1316, Kim \& Fabbiano 2003; Fig.~\ref{fig:1316distr}), 
 would be consistent with this hypothesis.
 However, in NGC4472 at least, no differences are found in the distributions of X-ray
 luminosities of the GC sources and the other LMXBs (Maccarone, Kundu \& Zepf 2003).
 Moreover, in the inner bulge of M31, at radii that even with {\it Chandra} cannot be explored
 in elliptical galaxies because of their distances, the distribution of LMXBs appears more
 peaked than that of the optical light (Kaaret 2002).
 
\subsection{X-ray Luminosity Functions \label{exlf}}

 The XLFs of the early-type galaxies observed 
with {\it Chandra} are generally steeper than those of star-forming galaxies (see Section~\ref{xlf}), 
i.e. with a relative lack of luminous
HMXBs. These XLFs are generally well fitted with power-laws or broken power laws
with (cumulative) slopes ranging from -1.0 to -1.8,
and breaks have been reported both at 2-3$\times 10^{38} ~\rm ergs~s^{-1}$, the
Eddington luminosity of an accreting neutron star (Sarazin, Irwin \& Bregman 2000; 
Blanton, Sarazin \& Irwin 2001; Finoguenov \& Jones 2002; Kundu, Maccarone  \& Zepf 2002), and 
at higher luminosities ($10^{39} ~\rm ergs~s^{-1}$) (Jeltema {\it et al.} 2003, in NGC~720).
While the former break may be related to a transition between neutron star and black hole
binaries (Sarazin, Irwin \& Bregman 2000), the latter, high luminosity break, 
could be produced by a decaying (aging) starburst component from binaries formed in
past merging and star bursting episodes (Wu 2001). This possibility was suggested in the case of
NGC~720 (Jeltema {\it et al.} 2003). The XLFs of NGC~5128 (Kraft {\it et al.} 2001),
obtained at different times and reflecting source variability, are  
well fitted with single power-laws in the  luminosity range of $10^{37} - 10^{39} ~\rm ergs~s^{-1}$.
In NGC~1291 (Irwin, Sarazin \& Bregman 2002), no super-Eddington sources are detected.

\begin{figure*}[htbp]
  \vbox{\hskip -\leftskip
  \psfig{figure=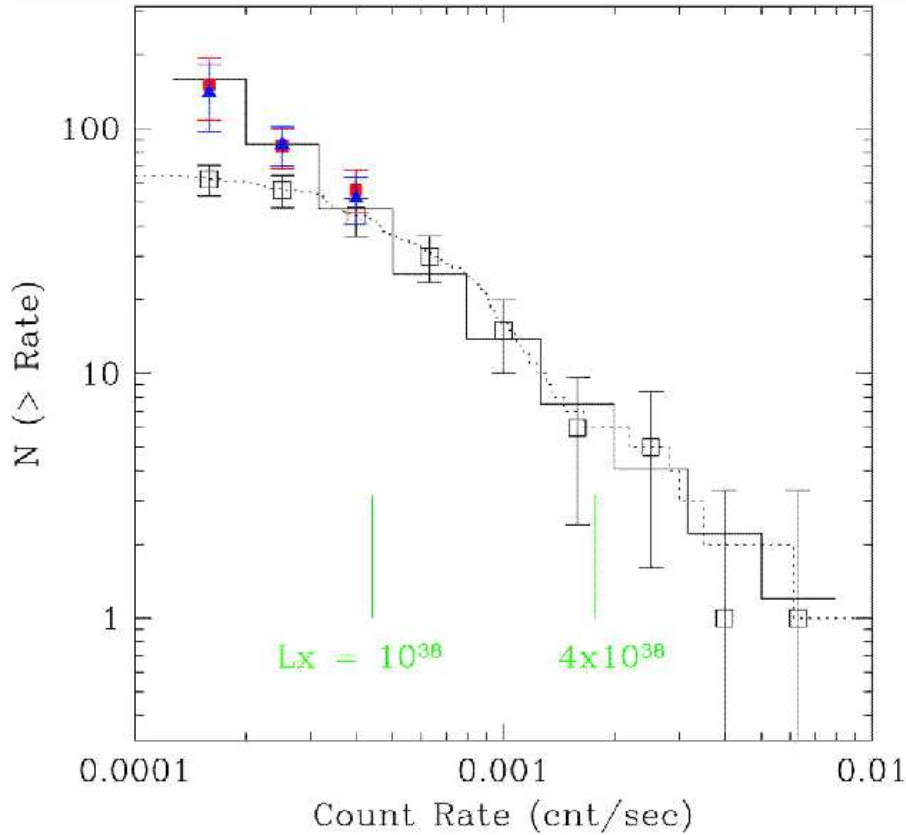,width=12.cm,angle=0.0}
  }\par
  %,bbllx=1.5cm,bblly=13.7cm,bburx=15.5cm,bbury=25.7cm,clip=}}\par
  \caption{\label{fig:1316XLF}
  Observed (empty squares) and corrected (filled points) XLFs of NGC~1316 (Kim \& Fabbiano (2003).}
\end{figure*}

The effects of detection incompleteness have been considered by Finoguenov \& Jones (2002),
and have been recently explored extensively by Kim \& Fabbiano (2003) in their derivation
of the XLF of NGC~1316. Low-luminosity sources may be missed because of higher background/diffuse
emission levels in the inner parts of galaxies, and also because of the widening of the
{\it Chandra} beam at larger radii. Correcting for these effects with an extensive set of simulations,
Kim \& Fabbiano (2003) found that an apparent 2-3$\times 10^{38} ~\rm ergs~s^{-1}$ break 
in the XLF of NGC~1316 disappeared when incompleteness was taken into account, and the 
XLF of this galaxy could be represented by an unbroken power-law down to  luminosities
of $\sim 3 \times 10^{37} ~\rm ergs~s^{-1}$ (Fig.~\ref{fig:1316XLF}). This result shows that caution must be exercised
in the derivation of XLFs, and that perhaps some of the previous reports should be reconsidered.
If the XLFs extend unbroken to lower luminosities, the amount of X-ray emission
from undetected LMXBs in early-type galaxies can be sizeable, as it is the case in NGC~1316.
This result is important not only for our understanding of the XRB populations, but also
for the derivations of the parameters of the hot interstellar medium in these system (see 
Kim \& Fabbiano 2003). Ignoring the contribution to the emission of hidden XRBs 
results in biases and erroneous results and may give the wrong picture of the overall
galaxy dynamics and evolution. Moreover, the dominance at large radii of XRB emission over
the hot ISM (see Fig.~\ref{fig:1316distr}) in some (X-ray faint) ellipticals, 
does also affect adversely mass measurements 
of these galaxies from low-resolution X-ray data (Kim \& Fabbiano 2003).

\section{Multi-wavelength Correlations \label{corr}}

Although this chapter is focussed on the XRB populations that we can now resolve
and study  with {\it Chandra} in galaxies as distant as $\sim 20$~Mpc, the study of
the integrated emission properties of samples of galaxies (either more distant,
or observed at lower resolution) can also give useful 
information on the average properties of their XRB components.
We will summarize here some of these studies, that were pursued mostly by
using the samples of galaxies observed with {\it Einstein} and {\it ROSAT}.

Most of the early work in this area was done by Fabbiano and collaborators, using
the first sample of galaxies ever observed in X-rays, the {\it Einstein} sample
(see reviews in Fabbiano 1989; 1995). Besides suggesting the baseline XRB emission
in E and S0 galaxy, that is now confirmed with {\it Chandra}  (Section~\ref{ell}),
these results suggested a general scaling of the integrated X-ray emission with
the optical luminosity (and therefore stellar population) of the galaxies, and
pointed to a strong association of the XRB populations of disk/arm-dominated
spirals with the far-IR emission, i.e. the younger component of the stellar population
(e.g., Fabbiano, Gioia \& Trinchieri 1988; see also David, Jones and Forman 1992). More recent 
work on the {\it Einstein} sample (Shapley, Fabbiano \& Eskridge 2001; Fabbiano \&
Shapley 2002\footnote{probably the last paper to be published on the {\it Einstein} data}),  
on {\it ROSAT}-observed galaxies (Read \& Ponman 2001), 
and on {\it Beppo-SAX} and {\it ASCA} data (Ramalli, Comastri \& Setti 2003) has
examined some of these correlations afresh.
Given the different pass-bands of these observatories, these studies have a varied
sensitivity to the effect of hard XRB emission and soft hot ISM emission in the galaxies.

The {\it Einstein} sample is the largest, consisting of 234 S0/a-Irr galaxies observed
in the 0.2-4.~keV band. The X-ray luminosities are compared with B, H, 12~$\mu$m, 60~$\mu$m,
100~$\mu$m, global FIR, and 6~cm luminosities (Shapley, Fabbiano \& Eskridge 2001;
Fabbiano \& Shapley 2002). 
Both fluxes and upper limits were used in this work, to
avoid obvious selection biases. This work provides baseline distributions of $L_X$ and of
$L_X/L_B$ for the entire Hubble sequence (including E and S0 galaxies), 
and a critical compilation of distances  for the sample. Multi-variable correlation
analysis shows clear dependencies of the emission properties on the morphological type of the galaxies
(and therefore indirectly on the stellar population and star formation activity).
In Sc-Irr galaxies, all the emission properties (including the X-rays) are tightly
correlated, suggesting a strong connection to the  stellar population. This is not
true for S0/a-Sab, where there is a general connection of the X-ray luminosity 
with the B and H-band emission (stellar population), but not with either radio or FIR.
In Sc-Irr galaxies the strongest link of the X-ray emission is a linear correlation with
the FIR, suggesting a connection with the star-forming stellar component.
This conclusion is reinforced by a correlation between $L_X/L_B$ and $L_{60}/L_{100}$,
which associates more intense X-ray emission with hotter IR colors.

The X-ray emission / star-formation connection is also discussed as a result of
the analysis of a small sample (17 nearby spirals) observed with ROSAT in a 
softer energy band (0.1-2.0~keV; Read \& Ponman 2001), and more recently
from the analysis of another small sample (also 17 galaxies)
observed in the 2-10~keV band
(Ranalli, Comastri \& Setti 2003). The advantage of this harder band is that
the emission is predominantly due to the XRB population (if the sample does not
include AGNs). These authors suggest that the hard X-ray emission can
be used as a clean indicator of star 
formation, because extinction is not a problem at these energies. 

These correlation analyses are now being extended to the XRB populations detected with
{\it Chandra}.
Colbert {\it et al.} (2003) report good correlations between the total point source
X-ray luminosity in a sample of 32 galaxies of different morphological type extracted 
from the {\it Chandra} archive and the stellar luminosity (both B and K bands).
While correlations are still present in the spiral and merger/irregular galaxies with FIR
and UV luminosities, the ellipticals do not follow this trend and show a clear lack of
FIR and UV emission, consistent with their older stellar populations. This results is
consistent with the conclusions of Fabbiano \& Shapley (2002; see above), which were however based
on the analysis of the integrated x-ray luminosity of bulge dominated and disk/arm 
dominated spiral and irregular galaxies.

In summary, there is a correlation between X-ray emission and SFR in star-forming 
galaxies, that may lead to a new indicator of the SFR. However, one has to exercise caution,
because this conclusion is only true
for star-forming galaxies. In old stellar systems (bulges, gas-poor E and S0s),
the X-ray emission is connected with the older stellar
population of these systems. This conclusion is also in agreement with the recent
studies of XLFs (Section~\ref{pops}; Section~\ref{exlf})

\section{The X-ray Evolution of Galaxies \label{deep}}

X-ray images of the extragalactic sky routinely taken with {\it Chandra} and
 {\it XMM-Newton} do not typically detect normal galaxies  as serendipitous 
 sources in the field. Instead the images reveal a relativity sparse 
 population of point sources, the majority of which are Active Galatic Nuclei (AGN) 
 with a space density of order a thousand per square degree. Normal galaxies are 
 not detected because the X-ray luminosity of normal galaxies is relatively low 
 and the predicted fluxes very faint. However, in the deepest few million 
 second or more exposures made with {\it Chandra} (the {\it Chandra} Deep Fields -- CDFs;
 Giacconi {\it et al.} 2002, Alexander \etal\ 2003) faint 
 X-ray emission has been detected from optically bright galaxies at redshifts 
 of 0.1 to 0.5  (Hornschemeier \etal\ 2001). 
 These are amongst the faintest X-ray sources in the CDF, with fluxes of 
 $\sim 10^{-16}$ \ergcms, corresponding to a luminosity of  $10^{39}$ to $10^{41}$ \ergs\ -- 
 the range seen from nearby galaxies (e.g. see Shapley, Fabbiano \& Eskridge 2001). 
 Some of these might be galaxies containing 
 a low luminosity AGN, but most are likely to be part of an emerging population of 
 normal galaxies at faint X-ray fluxes.  
 
 The detection sensitivity of {\it Chandra} can 
 be increased by `stacking' analysis, i. e. 
 by  `stacking'  sub-images centered on the positions of  galaxies 
 in comparable redshift ranges. This can push the threshold of {\it Chandra} to 
 $\sim 10^{-18}$ \ergcms\ -- equivalent to an effective exposure time of several 
 months or more. Brandt \etal\ (2001) used this technique for 24 Lyman Break galaxies 
 at z $\sim$ 3 in the Hubble Deep Field North (Steidel \etal\ 1996) and detected a 
 signal with an average luminosity of $3\times 10^{41}$ \ergs\ -- similar to that 
 of nearby starburst galaxies. Nandra \etal\ (2002) confirmed this result by 
 increasing the number of Lyman Break galaxies to 144 and then extended it to 
 also include 95 Balmer Break galaxies at $z \sim 1$. The Balmer Break galaxies 
 were  detected with a lower average luminosity of $7\times 10^{40}$ \ergs, but 
 a similar X-ray to optical luminosity ratio as the Lyman Break galaxies.
 Hornschemeier \etal\ (2002) report `stacking' detections of optically 
 luminous spiral galaxies at $0.4 < z < 1.5$.
 
 These {\it Chandra} X-ray Observatory detections of normal galaxies at high redshifts have 
 initiated the study of the X-ray evolution of normal galaxies over cosmologically 
 interesting distances. Evolution of the X-ray properties of galaxies is to be 
 expected because the star formation rate (SFR) of the universe was at least a 
 factor of 10 higher at redshifts of 1--3 (Madau \etal\ 1996). Since the X-ray 
 luminosity of galaxies scales with the infra-red and optical luminosity 
 (see Section~\ref{corr}; Fabbiano, Gioia \& Trinchieri 1988; 
 David, Jones and Forman 1992; Shapley, Fabbiano \& 
 Eskridge 2001; Fabbiano \& Shapley 2002) the increased star formation will have a corresponding  
 impact on the X-ray properties of galaxies at  high redshift (White and Ghosh 1998). 
 For spiral galaxies without an AGN, the overall X-ray luminosity in the 1--10 keV band 
 will typically be dominated by the galaxy's X-ray binary population. There is expected 
 to be a corresponding increase in the number of high mass X-ray binaries associated 
 with the increased star formation rate. The `detection' of the Balmer and Lyman Break 
 galaxies by Nanda \etal\ (2001) and the factor of 5 increase in the X-ray luminosity 
 from redshift 1 to 3 is consistent with an increasing star formation rate.
  Nandra \etal\ (2001) point out that the X-ray luminosity of galaxies provides 
  a new `dust free' method to estimate the star formation rate, as also pointed in the
  {\it Beppo-Sax} study of Ranalli, Comastri \& Setti (2003), and by Grimm, Gilfanov \&
  Sunyaev (2003).
  
The low mass X-ray 
  binary (LMXB) population created by the burst in star formation at z $>$ 1 may 
  not emerge as bright X-ray sources until several billion years later 
  (White and Ghosh 1998; Ghosh and White 2001).  This is due to the fact 
  that the evolutionary timescales of LMXBs, their progenitors, and their 
  descendants are thought be significant fractions of the time-interval between 
  the SFR peak and the present epoch.  In addition to an enhancement 
  near the peak ($z\approx 1.5$) of the SFR due to the prompt turn-on 
  of the relatively short-lived massive X-ray binaries, there may be a 
  second enhancement, by up to a factor $\sim 10$, at a redshift between 
  $\sim 0.5$ and $\sim 1$ due to the delayed turn-on of the LMXB 
  population (Ghosh and White 2001). This second enhancement will not be 
  associated with an overall increase in the optical or infrared luminosity 
  of the galaxy, resulting in an increase in the X-ray to optical luminosity ratio.  
  Hornschemeier {\it {\it et al.}} (2001) using the `stacking' technique detected 
  X-ray emission from $L_*$ redshift 0.4 to 1.5 spiral galaxies in the HDF-N. 
  The X-ray to optical luminosity ratios are consistent with those of galaxies 
  in the local universe (e.g., Shapley, Fabbiano \& Eskridge 2001),
   although the data indicate a possible increase in this 
  ratio by a factor of 2--3. 
  
\begin{figure*}[htbp]
  \vbox{\hskip -\leftskip
  \psfig{figure=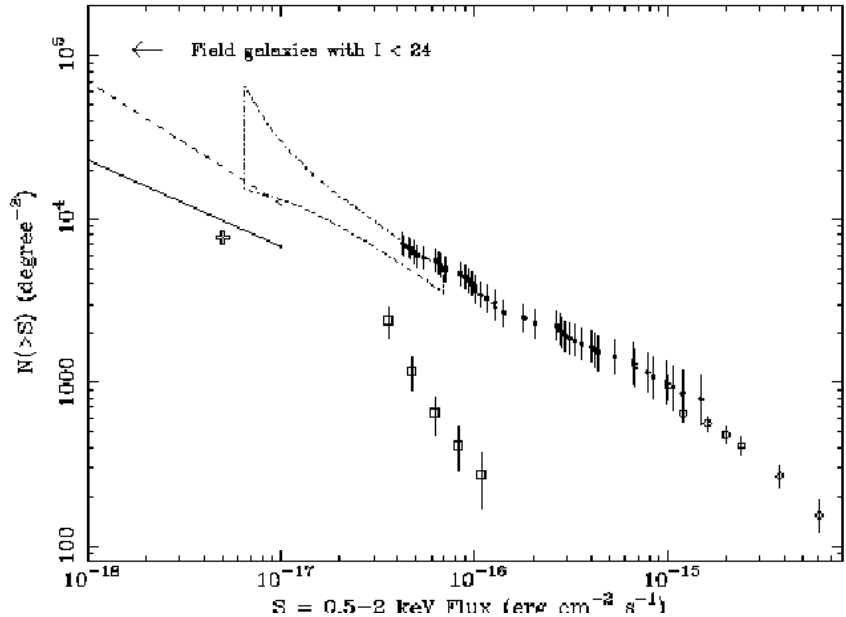,width=12.cm,angle=0.0}
  }\par
  %,bbllx=1.5cm,bblly=13.7cm,bburx=15.5cm,bbury=25.7cm,clip=}}\par
  \caption{\label{fig:deep}
  CDF-N number counts, with predictions (ar the faint end) based on different
  SRF at high redshift. The open boxes are the counts from the optically bright,
  X-ray faint sources - these are mainly normal and starburst galaxies, but some
  low-luminosity AGN may be present -. The cross is the result of the
  `stacking' analysis using $z  \leq 1.4$ galaxies in the CDF-N field. The solid and dashed 
  lines at the lowest fluxes are the the predictions of the galaxy number counts from
  Ptak et al 2001 (from Hornschemeier et al 2003). The leftward pointing arrow 
  indicates the number density of field galaxies at $I = 24$~mag.}
\end{figure*}

Ptak \etal\ (2001) discuss the observable consequences 
  of the increased SFR at high redshifts for the X-ray detection of galaxies at 
  redshift $>$ 1 in the HDF-N. To do this Ptak \etal\ (2001) used the Ghosh and White (2001) 
  models for the evolution of the underlying X-ray binary populations for several different 
  possible SFR models (the SFR with redshift is not well known). Depending on the 
  SFR model used, the average X-ray luminosity of galaxies in the HDF-N can be an 
  order of magnitude higher than in the local universe. These model predictions 
  can be translated into a prediction of the number counts verses flux. Fig.~\ref{fig:deep} 
  taken from Hornschemeier \etal\ (2003) shows the number counts from the CDF-N 
  (which are dominated by AGN), along with the predictions from Ptak \etal\ (2001) 
  for two different SFR models. The emerging population of optically bright, 
  X-ray faint (OBXF) galaxies detected in the CDF-N is also shown, along with 
  the extension of the source counts to fainter fluxes using a fluctuation analysis 
  of the CDF-N (Miyaji \& Griffiths 2002) . The predictions are that emission from 
  normal galaxies, largely at redshift of 1--3, will start to dominate the source 
  counts somewhere between fluxes of $10^{-17}$ and $10^{-18}$ \ergcms. The cross 
  on Fig.~\ref{fig:deep} shows the constraint from the stacking analysis of Hornschemeier 
  \etal\ (2002) for relatively nearby spiral galaxies (z$<$1.5), which is in 
  agreement with the predictions from Ptak \etal\ (2001) for the lower SFR models.

Much deeper {\it Chandra} exposures of several months or even a year long  will be able to 
  eventually reach fluxes of $10^{-18}$ \ergcms\  and directly test the models for 
  the X-ray evolution of galaxies --  given the projected long lifetime of {\it Chandra} 
  and good luck, these very deep exposures will hopefully will eventually happen as 
  the mission matures. To obtain spectra of these galaxies, which are typically at 
  a redshift of 1--3, and to see higher redshift objects at a similar faint flux level, 
  will require 100-1000 times more collecting area with 1 arc sec angular resolution 
  to avoid confusion (e.g., Fabbiano 1995, 2000; Elvis \& Fabbiano 1997;
  Fabbiano \& Kessler 2001). 
  Even more challenging, to resolve an AGN or an offset ULX from 
  more extended emission from the galaxy will require an angular resolution of 
  order 0.1 arc sec. Such mission parameters are technologically extremely challenging, 
  but nonetheless are being pursued by NASA, ESA and ISAS as a long term goal 
  for X-ray astronomy (Parmar \etal\ 2002, Zhang \etal\ 2002).

\section{Conclusions}

As we have shown in this review, X-ray studies of galaxies
are now yielding copious information on the properties of their XRB populations. The classification
and study of  these different populations is providing a unique tool for understanding the 
origin and evolution of XRBs, and for relating these sources to the evolution of the
stellar populations of the parent galaxies, both  in the nearby and  the far-away universe.

This work would not have happened without the vision of Riccardo Giacconi, who pushed 
forward the high resolution X-ray telescope concept, and the work of Leon Van Speybroeck,
who designed the {\it Chandra } optics.
We thank the colleagues that have provided figures and comments (Martin Elvis, Albert Kong, 
Ann Hornschemeier, Vicky Kalogera, Jon Miller, Doug Swartz, Andreas Zezas, Harvey Tananbaum,
Phil Kaaret, Jeff McClintock, Andrew King).
This work benefitted by the Aspen Summer Workshop on Compact X-ray Sources (Summer 2002),
and would not have been completed without the relentless prodding of Walter Lewin, to whom
we are indebted. We aknowledge partial support from the {\it Chandra} X-ray Center under
NASA contract NAS 8-39073.

\end{document}